\documentclass{aa}
\usepackage{graphicx}
\begin{document}
\title{Rotation and outflow in the central kiloparsec of the water-megamaser galaxies
IC\,2560, NGC\,1386, NGC\,1052, and Mrk\,1210\thanks{Based on observations collected at the
La Silla outstation of ESO. The data have been evaluated with the MIDAS data analysis system 
(version Nov. 96) provided by ESO}}
\author {H. Schulz\inst{1},
\and C. Henkel\inst{2} }
\offprints{H. Schulz, hschulz@astro.ruhr-uni-bochum.de}
\institute{Astronomisches Institut der Ruhr-Universit\"at, D-44780 Bochum, Germany
\and
Max-Planck-Institut f\"ur Radioastronomie, Auf dem H\"ugel 69,
D-53121 Bonn, Germany} 
\date{Received:FILL IN; accepted:FILL IN}
\authorrunning{Hartmut Schulz \& Christian Henkel}
\titlerunning{Rotation and outflow in water-megamaser galaxies}
\abstract{Optical emission-line profiles were evaluated in order to explore the 
structure of galactic nuclei containing H$_2$O megamaser sources. Long-slit spectra
of IC\,2560, NGC\,1386, NGC\,1052 and  Mrk\,1210 were obtained at $\sim 100$\,km/s 
spectral and $\sim 2\arcsec \times 2 \arcsec$ spatial resolution. The following 
individual properties of the objects were found: The active nucleus of IC\,2560 
(innermost $\pm 2\arcsec$) emits lines typical for a high-ionization Seyfert-2 
spectrum albeit with comparatively narrow profiles (FWHM $\sim 200$\,km/s). Line wings 
are stronger on the blue side than on the red side suggesting outflow. The central 
velocity gradient fits into the general velocity curve of the galaxy. Attributing 
it to a rotating disk coplanar with the galaxy leads to a Keplerian mass of $\sim 
10^7$\,M$_{\odot}$ inside a radius of 100 pc. --- The central 6\arcsec~sized structure seen 
on HST H$\alpha$ and [O{\sc iii}] images of NGC\,1386 appears to be the inner part of a 
near-edge-on warped rotating spiral disk that is traced in H$\alpha$ within a diameter 
of 17\arcsec~along p.a.\,23\degr. This interpretation is based on observed kinematic 
continuity and a typical S-shaped dust lane crossing the kinematical center. The 
central velocity gradient yields a Keplerian mass estimate of $\sim$10$^8$\,M$_{\odot}$ inside
$R=100$ pc and 5\,10$^9$\,M$_{\odot}$ inside 0.8 kpc. The total mass of the ionized gas
of $10^{5-6}$\,M$_{\odot}$ is small compared to the dynamical mass of the spiral disk. 
--- The kinematical gradient of the rotating gas disk in the center of the elliptical 
galaxy NGC\,1052 yields a mass of $\sim$6\,10$^8$\,M$_{\odot}$ inside 166\,pc. Two 
components are found: component $\cal A$ arises in a rotating disk, component 
$\cal B$ is blueshifted by $\sim400$\,km/s relative to the disk. $\cal B$ likely 
originates from outflowing gas distributed within a wide cone. There is no need for 
a broad H$\alpha$ component in unpolarized flux. The peculiar polarization seen 
in [O{\sc i}]$\lambda 6300$ by Barth et al. (1999) may be related to the outflow 
component. --- In Mrk\,1210, a redshifted component $\cal R$ contributes to the 
exceptional width of [O{\sc i}]$\lambda 6300$. An additional blueshifted component $\cal B$ 
strong in [O{\sc iii}] allows to fit the H$\alpha$ + [N{\sc ii}] blend without a broad component 
of H$\alpha$. $\cal B$ and $\cal R$ are likely to be outflow components. The location 
of the brightness maximum BM is slightly ($\sim1\arcsec$) shifted relative to the 
kinematical center KC. Locating the true obscured nucleus in KC makes it understandable 
that BM displays faint broad scattered H$\alpha$ in polarized light. --- Galactic 
rotation and outflow of narrow-line gas are common features of this sample of 
water-megamaser galaxies. All decomposed line-systems exhibit AGN typical line ratios. 
Recent detections of H$_2$O megamasers in starburst galaxies and the apparent 
asssociation of one megamaser with a Seyfert 1 AGN suggest that megamasers can possibly 
be triggered by optically detectable outflows. The frequently encountered edge-on geometry 
favoring large molecular column densities appears to be verified for NGC\,1386 and 
IC\,2560. For NGC\,1052 and Mrk\,1210, maser emission triggered by the optically
detected outflow components cannot be ruled out.  
}

\maketitle

\section{Introduction}
During the past two decades, powerful H$_2$O masers have been detected in the 
6$_{16}$--5$_{23}$ transition at 22\,GHz ($\lambda$$\sim$1.3\,cm) towards almost
thirty galaxies known to contain an active galactic nucleus (AGN) of Seyfert 
2 or LINER type (dos Santos \& L{\'e}pine 1979; Gardner \& Whiteoak 1982; Claussen 
et al. 1984; Henkel et al. 1984, 2002; Haschick \& Baan 1985; Koekemoer et al. 1995; 
Braatz et al. 1994, 1996, 1997; Greenhill et al. 1997a, 2002, 2003; Hagiwara et al. 1997; 
Falcke et al. 2000). These H$_2$O masers (hereafter `megamasers') appear to be more 
luminous, by orders of magnitude, then the brightest H$_2$O masers observed in galactic 
star forming regions. All interferometric data of H$_2$O megamasers point toward a 
nuclear origin. The emission originates from the innermost parsec(s) of the parent 
galaxy (e.g. Claussen \& Lo 1986; Greenhill et al. 1995b, 1996, 1997b; Miyoshi et al. 1995; 
Claussen et al. 1998; Trotter et al. 1998; Herrnstein et al. 1999). Adopting the 
so-called unified model (Antonucci \& Miller 1985; Antonucci 1993) with Seyfert 
2 nuclear disks being viewed edge-on, megamaser activity must be related to large 
column densities along the line of sight, a picture in line with the presence of 
dense molecular gas. 

There exist at least two different classes of megamasers, those forming a 
circumnuclear disk (like in NGC\,4258) and those associated with nuclear jets 
(e.g. NGC\,1052). The observed H$_2$O transition traces a dense ($\ga$10$^{7}$\,cm$^{-3}$) 
warm ($T_{\rm kin}$$\ga$400\,K) molecular medium. The additional presence of jets and 
a nuclear X-ray source requires, however, the coexistence of neutral and ionized gas 
in the nuclear regions of these galaxies. In an attempt to study connections between
megamaser emission on milliarcsecond scales and optical phenomena on arcsecond scales, 
here we present and analyze optical emission line spectra from four megamaser galaxies.
\begin{table*}
\begin{minipage}{180mm}
\caption[]{Journal of the spectroscopic observations}
\begin{flushleft}
\begin{tabular}{lllcllc}
\hline
Object/helioc. corr.\footnote{This heliocentric correction was added to the measured 
 radial velocities.} & Frame & Date (beg.) & p.a. (slit) & Pos. of slit &
 $\lambda$ range (\AA) & Exp.time (s) \\
\hline\hline
NGC\,1052&F020&11-Jan-96&90\degr&through nucleus& 5400-7400 & 3600\\
$-27$ km/s&F056 & 12-Jan-96 & 90\degr & through nucleus& 5400-7400 & 1800\\
&F057 & 12-Jan-96 & 90\degr & 2\arcsec~south of nucleus & 5400-7400 & 1800\\
&F058 & 12-Jan-96 & 90\degr & 2\arcsec~north of nucleus & 5400-7400 & 1800\\
&F078 & 13-Jan-96 & 45\degr & through nucleus & 5000-6800 & 3600\\
&F107 & 14-Jan-96 & 45\degr & through nucleus & 4000-6000 & 3600\\
&F108 & 14-Jan-96 & 45\degr & 2\arcsec~northwest of nucleus & 4000-6000 & 3600\\
\hline
NGC\,1368& F023 & 11-Jan-96 & 90\degr & through nucleus & 5400-7400 & 3600 \\
$-17$ km/s & F026 & 11-Jan-96 & 90\degr & 2\arcsec~north of nucleus & 5400-7400 & 3600 \\
&F080 & 13-Jan-96 & 23\degr & through nucleus & 5000-6800 & 2700\\
&F112 & 14-Jan-96 & 23\degr & through nucleus & 4000-6000 & 3600\\
\hline
Mrk\,1210 & F066 & 12-Jan-96 & 90\degr & through nucleus & 5400-7400 & 3600 \\
$+6$ km/s &F084 & 13-Jan-96 & 90\degr & through nucleus & 5000-6800 & 1333\\
&F085 & 13-Jan-96 & 90\degr & 2\arcsec~south of nucleus & 5000-6800 & 3600\\
&F087 & 13-Jan-96 & 90\degr & 2\arcsec~north of nucleus & 5000-6800 & 3600\\
&F116 & 14-Jan-96 & 90\degr & through nucleus & 4000-6000 & 3600\\
\hline
IC\,2560 & F030 & 11-Jan-96 & 90\degr & through nucleus & 5400-7400 & 3600 \\
$+20$ km/s & F118 & 14-Jan-96 & 90\degr & through nucleus & 4000-6000 & 3600 \\
\hline
\end{tabular}
\end{flushleft}
\end{minipage}
\end{table*}
\section{Data acquisition}
The spectra were obtained in January 1996 using the Boller \& Chivens spectrograph 
attached to the Cassegrain focus of the ESO 1.52\,m spectrographic telescope. 
A log of the observations, in particular the observed wavelength ranges, are given 
in Table 1. The detector was La Silla CCD No. 24 (FA2048L CCD chip with 15\,$\mu$m 
wide square pixels). The spatial resolution element is 0\farcs68 px$^{-1}$. Seeing 
and telescope properties limited the spatial resolution to the range 1\farcs2 -- 
2\farcs5 which was determined by the width of standard star spectra on the focal
exposures. The slit was visually centered and guided on the nuclear brightness maximum 
or a nearby offset star. The 2\arcsec ~wide slit projects to a spectral resolution 
of $\sim 2$\AA~ ($\sim 100$\,km s$^{-1}$) as is confirmed by the full width at 
half-maximum (FWHM) of comparison lines and the [O{\sc i}]$\lambda6300$ night sky line.

Employing the ESO MIDAS standard software (version Nov. 96) the CCD frames were bias 
subtracted and divided by well exposed flat-field frames. Night-sky spectra `below' 
and `above' any notable galaxy emission were interpolated in the region of the galactic 
spectrum and subtracted in each case. From such cleaned frames either single rows 
(corresponding to 0\farcs68 and 2\arcsec~along and perpendicular to the slit direction, 
respectively) or the averages of two adjacent spectral rows ($1\farcs36 \times 2\arcsec$ 
formal spatial resolution) were extracted. 
 
T\"ug's (1977) curve was used to correct for atmospheric extinction. The spectra were 
flux calibrated using the standard star 40 Eri B (Oke 1974), but we use the calibration 
only for relative line ratios due to occasional thin cirrus. Forbidden-line wavelengths
were taken from Bowen (1960). Spectra in the blue and red region of the bright elliptical 
NGC\,3115 were taken as well and we used them as templates to subtract the stellar 
contribution from those spectra in which we analyzed the line-profile structure and 
intensity ratios.

\section{General data and morphology}
Table 2 (column 2) exhibits the morphological diversity of the four galaxies although we only
refer to the crude Hubble type. While NGC\,1052 is an elliptical galaxy (Fosbury et al.\,1978), 
NGC\,1386 (Sa according to Weaver et al.\,1991) and IC\,2560 (Sb in Fairall 1986; SBb in 
Tsvetanov \& Petrosian 1995) are disk systems whereas Mrk\,1210 is hard to classify from 
ground-based data. It looks roundish with an amorphous body supplemented by a few 
shell-like features, possibly a late-stage merger (Heisler \& Vader 1994). An HST-snapshot 
image unveils a kpc-sized central spiral apparently seen close to face-on that can be 
classified as Sa in a Hubble scheme (Malkan et al. 1998), although it should be cautioned 
that this spiral may be gaseous rather than a density-wave triggered stellar spiral. 

Column 3 gives the heliocentric value of the systemic velocity as derived from our data.
All velocity values given below are heliocentric. Column 4 of Table\,\ref{t2} gives distances 
(see footnotes) in accordance with recent work to ease comparison. Column 5 gives the linear 
scale in the galaxy applicable along the major axis for the distance (col. 4) assumed. 
Cols. 6 and 7 give the position angle of the apparent major axis and the angle of inclination 
from face-on. For the present work, these numbers should preferably correspond to the disk 
in which emission lines are generated, which is, however, only poorly determined. Comments 
and footnotes explain the origin of the numbers given.  

\section{Results on individual galaxies}
Details of the spectral data analysis to find clues for basic kinematical bulk components
in the shape of the emission lines are outlined in Appendices A -- B (treatment of individual 
spectra) and Appendix C (determination of rotation curves). Here we discuss our results, 
starting with the two strongly inclined spiral galaxies IC\,2560 and NGC\,1386. Then we 
turn to the apparently more complex spectra of NGC\,1052 and Mrk\,1210. 
\subsection{The Seyfert IC\,2560}
\label{ic25}
\begin{table*}
\begin{minipage}{180mm}
\caption[]{Adopted basic data of the four galaxies ($v_{\rm hel}$ from this work)}
\label{t2}
\begin{flushleft}
\begin{tabular}{llllllll}
\hline
Object & Type & $v_{\rm hel}$ &  dist. & lin. sc. & maj. ax. &
inclin.  &
 Comment \\
Coord.(1950) & Hubble & km s$^{-1}$ & Mpc & pc/\arcsec& \degr & \degr & \\
\hline\hline
NGC\,1052 & E & $1462\pm20$ &  17.1\footnote{distance from $v_{\rm recess}=1282$\,km/s relative to CMBR 
(de Vaucouleurs et al.\,1991) as used in Claussen et al. (1998)} & 82.9 & 49\footnote{Davies \& 
Illingworth (1986)}  & $\sim 60$\footnote{from this work} & p.a. (major axis) and inclination \\
0238-08 &    &    &    &    &    
&     & of nuclear gas disk \\
\hline
NGC\,1386 & Sa & $878\pm20$ & 19.8\footnote{we assume the same distance as for NGC\,1365 in Fornax
cluster;
cf. Schulz et al. (1999)} & 96.0 & 23 & 74 & major axis and inclination of galaxy\\
0334-36     &    &    &      &     &     &  & on large scale (Tully 1988) \\  
\hline
Mrk\,1210 & amorph & $4022\pm30$ & 53.6\footnote{distance via direct application of
$H_{\rm 0}=75$\,km s$^{-1}$ Mpc$^{-1}$} & 260.0 & $\approx 110$\footnote{kinematical major axis, this work; see
Sect.\,\ref{fou}} & $\approx 0 $   
& maj. ax. from gas kinematics; inclin. from image\\
0801+05     &
 Sa\footnote{type of inner kpc-spiral; based on HST image (Malkan et al. 1998)}   &  &  &   & & & 
of kiloparsec spiral (Malkan et al.\, 1998) \\
\hline
IC\,2560  & SBb   & $2922\pm20$\footnote{$v_{\rm hel}$ measured by us at the location of maximal brightness; 
the value agrees with that of HI (Mathewson et al. 1992)}    & 26\footnote{member Antlia cluster; 
distance from Aaronson et al. (1989) as used in Ishihara et al. (2001)}   & 126.1    &
  45   & 63\footnote{major axis and inclination of galaxy on 
large scale (Mathewson et al. 1992)}   & p.a. and inclination of galaxy; distance \\
1014-33   &     &    &    &     
&     &    & 39 Mpc if peculiar motion is ignored \\
\hline
\end{tabular}
\end{flushleft}
\end{minipage}
\end{table*}
\subsubsection{Kinematics} 
\begin{figure}[ht]
\resizebox{\hsize}{!}{\includegraphics[angle=-90]{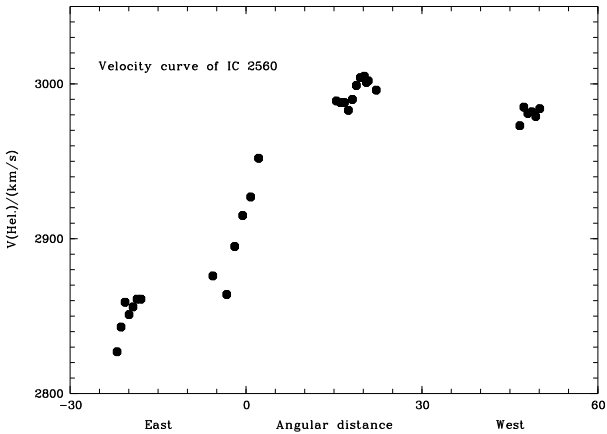}}
\caption{Heliocentric radial velocities (averages of the peaks of H$\alpha$ and [N{\sc ii}]$\lambda6583$)
measured in spectrum F030 of IC\,2560 (along p.a. 90\degr; abscissa in arcsec). Statistical 
uncertainty $\pm 11$\,km/s. } 
\label{vic}
\end{figure}

Along p.a.\,(90\degr, 270\degr) we traced regions with emission lines 22\arcsec~east 
until 50\arcsec~west. The reference position 0\arcsec~on the abscissa of Fig.\,1 marks the 
spot where the maximal intensity of nuclear line emission is reached. Here 
$v_{\rm hel}^{\rm c}=(2922\pm 20$)\,km/s is measured, which we identify with the systemic 
velocity.  Outside the innermost nuclear region (useful line measurements are possible 
from the nuclear maximum within a diameter of 5\arcsec) appears a gap with no traces of 
a line and emission lines further out originate in disk HII regions. The corresponding 
velocity curve is shown in Fig.\,\ref{vic}. Between 15\arcsec~and 20\arcsec~west
an interesting local up- and downturn (presumably density-wave related) can be noted 
in the spiral arm. This and two other intersections of a spiral arm ($\sim$ 45\arcsec~to 
50\arcsec~west and 17\arcsec~to 22\arcsec~east) exhibit contiguous H$\alpha$ emission 
and mostly [N{\sc ii}] as well, with each region a few hundred parsec in extent. 
 
To evaluate the central linear gradient a deprojection onto the plane of the galaxy is 
required. Here $\theta=45\degr$~is the angle in the plane of the sky between slit position 
and major axis, because the slit extends along p.a.\,90\degr~and the galaxy's major axis 
along 45\degr, while its inclination $i=63\degr$~(Table \ref{t2}). 

Altogether, a nuclear gradient in rotational velocity of $20$\,km/s/(100 pc) is obtained 
yielding a Keplerian mass estimate ($M=R v^2/G$) of $M_{100}$ = 9.3\,10$^6$\,M$_{\odot}$ within 
a nuclear radius $R=100$ pc. This is only 2.6 times the value that Ishihara et al. (2001) 
obtain for the mass inside the central 0.07 pc (note that they assumed an edge-on 
disk). Extrapolating our gradient to this radius yields a negligible velocity compared to 
the actual velocities of the maser sources (210--420\,km/s) measured by Ishihara et al. (2001)
between 0.07 and 0.26 pc. This might be explained by an inner Keplerian upturn of the 
rotation curve due to the central mass concentration. 

\begin{figure}[ht]
\resizebox{\hsize}{!}{\includegraphics[angle=-90]{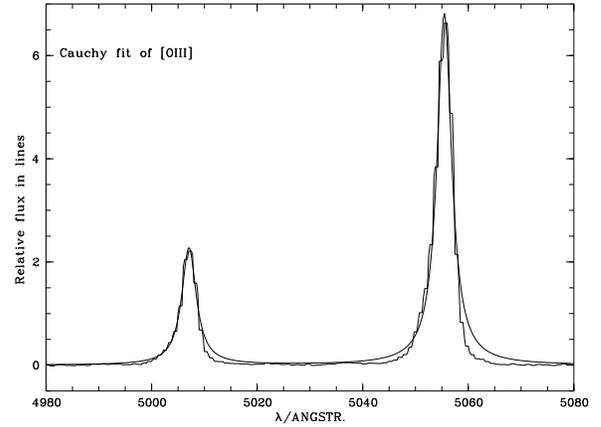}}
\caption{[O{\sc iii}]$\lambda\lambda4959, 5007$ from the central 1\farcs36 (E-W) $\times$ 2\arcsec (N-S) 
of IC\,2560 (F118) (plotted in the observed frame as all subsequent spectra). The Lorentzians used 
for the fits are too wide within the red wings.}
\label{cau}
\end{figure}

\begin{figure}[ht]
\resizebox{\hsize}{!}{\includegraphics[angle=-90]{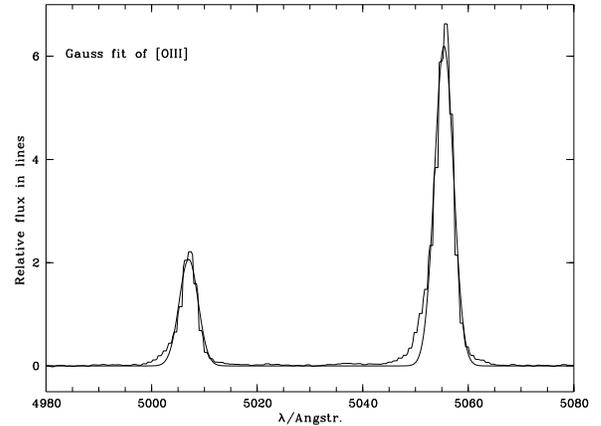}}
\caption{The same [O{\sc iii}]$\lambda\lambda4959, 5007$ observations as in Fig.\,2, but now fit by Gaussians, 
which turn out to be too narrow to fit the blue wings.}
\label{gau}
\end{figure}

The line profiles from the nuclear region within $\pm 2\arcsec$~of the center (defined by the 
maximum of continuum and line brightness) are relatively narrow for a Seyfert, with FWHMs 
of (200--250)\,km/s. Fig.\,\ref{cau} shows that the blue side of the [O{\sc iii}] profiles can be 
fit by a Lorentzian while Fig.\,\ref{gau} shows that the red side of the [O{\sc iii}] profiles 
can be fit by a Gaussian (this also holds for H$\beta$, H$\alpha$ and [N{\sc ii}]). This indicates 
that the observed [O{\sc iii}] profiles have a stronger outer wing on the blue side, which is most 
likely a signature for outflow. However, the amount and velocity of the outflow cannot be 
modelled precisely because the line width is only twice our spectral resolution. Taking the 
Gaussian as representative for the rotational component convolved with the instrumental profile,
the deconvolved l.o.s. outflow-velocity component is estimated to be $\sim 100$\,km/s. 

It should be noted that our near-edge-on view of the central kpc-disk may be unfavorable to 
measure outflowing gas, because outflow is frequently directed perpendicular to a galactic disk.
%
%
%
%
%
%
%
%
%
%
%

\subsubsection{Trends in line-ratios}
For readers interested in line-excitation variations a few trends of line ratios are given in the 
present, albeit primarily kinematic, study. In the innermost $\pm 2\arcsec$~ along p.a. (90\degr, 
270\degr) the intensity ratios of [N{\sc ii}]/H$\alpha$ and [S{\sc ii}]/H$\alpha$ are increasing to the west. 
At the three positions, central maximum, 1\arcsec~west, and 2\arcsec~west, we measure: 
[O{\sc i}]$\lambda6300$/H$\alpha =$ 0.15, 0.2 and `not measurable'; [N{\sc ii}]$\lambda6583/$H$\alpha =$ 
1.05, 1.2 and 1.5; [S{\sc ii}]($\lambda6716+\lambda6731)/$H$\alpha =$ 0.6, 0.65 and 0.9, respectively.
[O{\sc iii}] was extracted over 2\arcsec~bins yielding [O{\sc iii}]$\lambda5007/$H$\beta = 13$ in the central 
bin, which decreases to 10.5 at 2\arcsec~west, while it may slightly increase to 14 at 2\arcsec~east.
The other line ratios at 1\arcsec~east and 2\arcsec~east do not significantly differ from the
central values. The density-indicator [S{\sc ii}]$\lambda6716$/[S{\sc ii}]$\lambda6731$ is found in the range 
$0.90 \pm 0.05$ in this inner region, corresponding to single-component densities in the range 
$(7-10)\,10^2$ cm$^{-3}$. 

The relative strengthening of the low-ionization lines to the west (which is also the far side
if the spiral is trailing) may be due to a decrease of the ionization parameter by increased filtering
of the ionizing radiation with nuclear distance. Such an effect is demonstrated in Schulz \& Fritsch 
(1994). The directly measured line intensities of H$\alpha$ have a steeper fall-off to the west
than to the east. 

Exploratory spectra with an east-west oriented slit positioned 2\arcsec, 4\arcsec~and 6\arcsec~south 
of the nucleus (taken to look for some circumnuclear influence of the activity) are too weak for 
detailed measurements. At 4\arcsec~south the spectrum appears to be of lower excitation than known 
from the standard LINER sample shown in the diagnostic diagrams of Schulz \& Fritsch (1994).
%
%
%
%
\subsection{The Seyfert NGC\,1386}
\label{n13}
\subsubsection{Global kinematics of the inner emission-line region}
\label{glo13}
\begin{figure}[ht]
\resizebox{\hsize}{!}{\includegraphics[angle=-90]{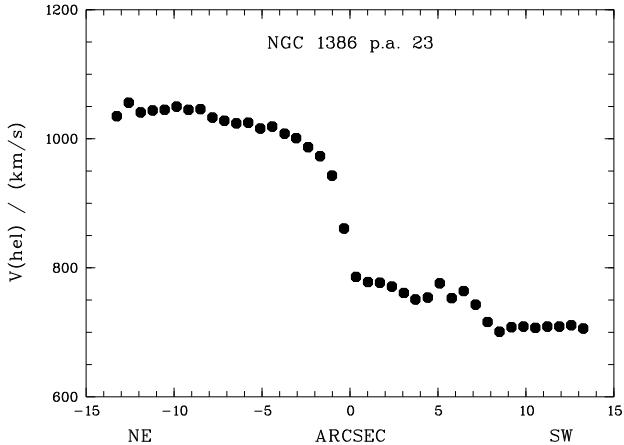}}
\caption{Velocity curve of NGC\,1386 along p.a. 23\degr (from F080). The data correspond to the 
mean position of the line peaks of H$\alpha$ and [N{\sc ii}]$\lambda6583$ in the raw spectrum.}
\label{v13}
\end{figure}

We here first describe a velocity curve derived from the line peaks of H$\alpha$ and 
[N{\sc ii}]$\lambda6583$ measured along the large-scale major axis at p.a. 23\degr~(Fig.\,\ref{v13}). 
The [O{\sc iii}]$\lambda\lambda4959, 5007$ (spectrum F112) peak positions agree in the innermost $\pm 
3\arcsec$ where they can be measured. For a comparison of the complicated line profiles of
[O{\sc iii}] and H$\alpha$ see Sect.\,\ref{in13}. H$\beta$ is faint, partially due to strong extinction 
(Sect.\,\ref{in13}), and cannot be measured beyond $\sim 2\farcs5$~of the brightness maximum. 
%
%

The zero point of the abscissa of Fig.\,\ref{v13} corresponds to the spectral row where continuum
and intensities of H$\alpha$ and [O{\sc iii}] are at maximum. We call this brightness maximum `region 
$\cal{B}$' because of its blueshift\footnote{Blueshifted regions in NGC\,1052 and Mrk\,1210 will be 
called $\cal{B}$ as well} relative to the kinematical center KC (see Fig.\,\ref{v13}). KC is defined 
by taking the symmetry center of the outer portions of the velocity curve. For its determination we took
means of the flat portions of the velocity curve beyond $\pm 8\arcsec$~ yielding (after symmetrizing) 
$v=168$\,km/s (northeast) and $v=-168$\,km/s (southwest) around $v_{\rm KC} = 877$\,km/s. Below in 
Sect.\,\ref{in13} we will find that the mean of the line cores of six different lines arising within 
the innermost 2\arcsec~of KC yields 878\,km/s, which is in (fortuitously) perfect agreement with 
$v_{\rm KC}$. 

Linearly interpolating in the spatial direction, KC is found to lie $-0\farcs6$~(to the northeast) 
offset from $\cal{B}$. 

The position of $\cal{B}$ on the sky can be inferred from an HST H$\alpha$~image shown in Ferruit et 
al. (2000) (their Fig.\,4, left panel, middle row; hereafter FWM2000). The following description 
refers to their H$\alpha$ image. Our spectrograph's slit encompasses the bright parts of the image 
by a 2\arcsec~wide band along p.a. 23\degr. The major axis of the H$\alpha$ emission string appears 
to extend along p.a. 6\degr, or $\Delta\theta = 17\degr$ offset from the slit position angle. 

From the rotation pattern and the appearance of NGC\,1386 as a highly inclined galaxy one expects 
to look upon a similarly inclined emission-line disk in the center. Images of galaxies at high 
inclination commonly show a midplane-dust lane. In FWM2000's H$\alpha$ image appears indeed a dark 
lane crossing the center between the two brightest spots that seems to wind from the southwest to 
the northeast in a extended S-shaped pattern.

The {\em dominating spot} in FWM2000's image, just below the dust lane, has to coincide with 
$\cal{B}$. $\cal{B}$ is located at the zero position in our Fig.\,\ref{v13}. The kinematical 
center KC, offset by 0\farcs6 in our spectrum, has to be located within the dust lane.

The H$\alpha$ image structure is characteristic for a near-to-edge-on and warped two-armed spiral disk. 
The spiral bending is in the same sense as that of the large galaxy, but the inclination appears to 
be larger than $i = 74\degr$, the inclination of the large scale disk (Table 2). Since the 
northeastern part is receding, the northwestern arm should be on the near side if we make the usual 
assumption that the arms are trailing. 

The above interpretation of the main extended structure in FWM2000's H$\alpha$ image 
may be considered to be at variance with the usual notion of a `radiation
cone along the minor axis'. When interpreting IR spectroscopy, Reunanen et al. (2002) use 
the term `radiation cone' to describe the morphology of the emission-line gas, but, on the other hand, 
they note that the H$_2$ ``rotation curve is relatively ordered parallel to the cone, suggesting that 
circular motions dominate the kinematics''. Our data agree with this statement on the kinematics,
but it is not clear whether the edge-on disk is ionized by a UV `radiation cone' (small plumes
along p.a. $90^{\circ}$ could be due to outflowing gas illuminated by a radiation cone; see
end of Sect.\,4.2.2).

The interpretation in terms of dominating circular motions of the inner ionized gas is further supported by
kinematical continuity with gas at larger galactocentric distances.
The bright H$\alpha$ features seen between 4\arcsec~northeast and 2\arcsec~southwest in the HST image 
are kinematically smoothly joined with gas further out (that is traced in our spectra but not 
in the HST image) to 14\arcsec~northeast and to 14\arcsec~southwest (Fig.\,\ref{v13}). Furthermore, 
the map of peak velocities in Weaver et al. (1991; their Fig.\,5) shows a slightly distorted, but 
nonetheless typical pattern of a rotating disk. An interpretation by a bipolar flow accidentally 
aligned with the kinematical major axis of the rotating disk would appear contrived, because it had to be
essentially decoupled from the gravitational field and its accordingly stratified interstellar medium. 

Altogether, the natural explanation of our Fig.\,\ref{v13} is that of a slightly projected rotation 
curve, although the rotating rings may not have a common plane, because FWM2000 noted changes of 
ellipticity and p.a.\,(major axis) with galactocentric radius. Owing to its AGN-typical line ratios
the H$\alpha$ gas is ionized by nuclear AGN radiation, but since the disk is likely to be matter-bounded
the UV radiation is not necessarily confined to a narrow cone. Therefore we dismiss the term
`radiation cone' for NGC\,1386.

For estimates of the rotation velocity utilizing Eq.\,\ref{rot} we have to know the orientation of 
the supposed {\em single} disk. Regarding apparent distances a projection correction would be small 
(because $\cos 17\degr = 0.96$). A velocity correction with $i=74\degr$~and $\theta = 17\degr$~would 
formally yield a factor 1.5 to obtain a rotational velocity. However, this procedure is questionable 
because $i=74\degr$~corresponds to the outer galactic disk and the inner disk addressed here appears 
to be more inclined.

For Keplerian mass estimates (that are good to a factor of two even if the potential is flattened; cf. 
Lequeux 1983) it is therefore sufficient to assume that the slit extends along the major axis and that 
the disk is seen edge-on, i.e. measured velocities are rotational velocities and angular distances 
along the slit correspond directly to linear distances (an inclination correction solely with $\sin 
74\degr$~would lead to 8\% larger masses).

The HST image (FWM2000's Fig.\,4) only shows the inner 6\arcsec~of the disk where H$\alpha$ is brightest.
As noted above, the velocity curve in our Fig.\,\ref{v13} extends into the galactic disk beyond the bright
H$\alpha$ clouds seen in the HST image. The curve becomes flat at $\pm 8\farcs5$~corresponding to a 
galactocentric radius $R = 816$ pc at which $v=168$\,km/s. These numbers yield a mass of $M = R v^2/G = 
5.4\,10^9$\,M$_{\odot}$~in the central region of 1.6 kpc diameter of NGC\,1386. In the center, the steep 
rise of the rotation curve to $\sim 90$\,km/s at 1\arcsec~yields a mass of $1.8\,10^8$\,M$_{\odot}$ within 
a diameter of 200 pc. 

From the observed H$\alpha$ luminosity $L_{{\rm H}\alpha} \approx 10^{40}$ erg/s (Colbert et al. 1997)
and adopting $L_{{\rm H}\alpha} = 2.9\  L_{{\rm H}\beta} $~ one can estimate the volume emission 
measure $V n_e^2$  for fully ionized ($n_e=n_p$, the proton density) hydrogen gas via $L_{{\rm H}\beta} 
= (V n_e^2)(4\pi j_{{\rm H}\beta}/n_e^2)$~($j_{{\rm H}\beta}$ is the emission coefficient of H$\beta$ 
per steradian).  Taking from Osterbrock (1989; Table\,4.2) $4\pi j_{{\rm H}\beta}/(n_e n_p) = 
1.24\,10^{-25}$ erg cm$^3$ s$^{-1}$ (for $T=10^4$ K) and adopting a typical NLR electron-density range 
$n_e = 10^{2-3}$ cm$^{-3}$, applying a conservative extinction correction of a factor of 3 and a factor 
of 1.45 for the presence of helium, one obtains $10^{5-6}$\,M$_{\odot}$ for the mass of line-emitting gas, 
which is small compared to the dynamical masses given above.  

\subsubsection{Line profiles from the inner disk}
\label{in13}
For sampling in closer accordance with the seeing we binned pairs of the 0\farcs68 wide CCD rows to 
spectral scans, which each encompass light from 1\farcs36 along the entrance slit, and subtracted 
scaled spectra of our template NGC\,3115 to remove the contribution of stellar light. H$\beta$ turned 
out to be faint and too noisy for an {\em accurate} estimate of the extinction from the Balmer 
decrement. Roughly, at the position of the line maximum we have $I$(H$\alpha$)/$I$(H$\beta$) $\approx 
4$, while this value approaches 6 at 1\farcs36 southwest, and 7 at 1\farcs36 northeast. In a simple 
screen model these values translate into $A_{\rm V} = 0\fm9 $, $2^m$~and 2\fm5, respectively.  

\begin{figure}[ht]
\resizebox{\hsize}{!}{\includegraphics[angle=-90]{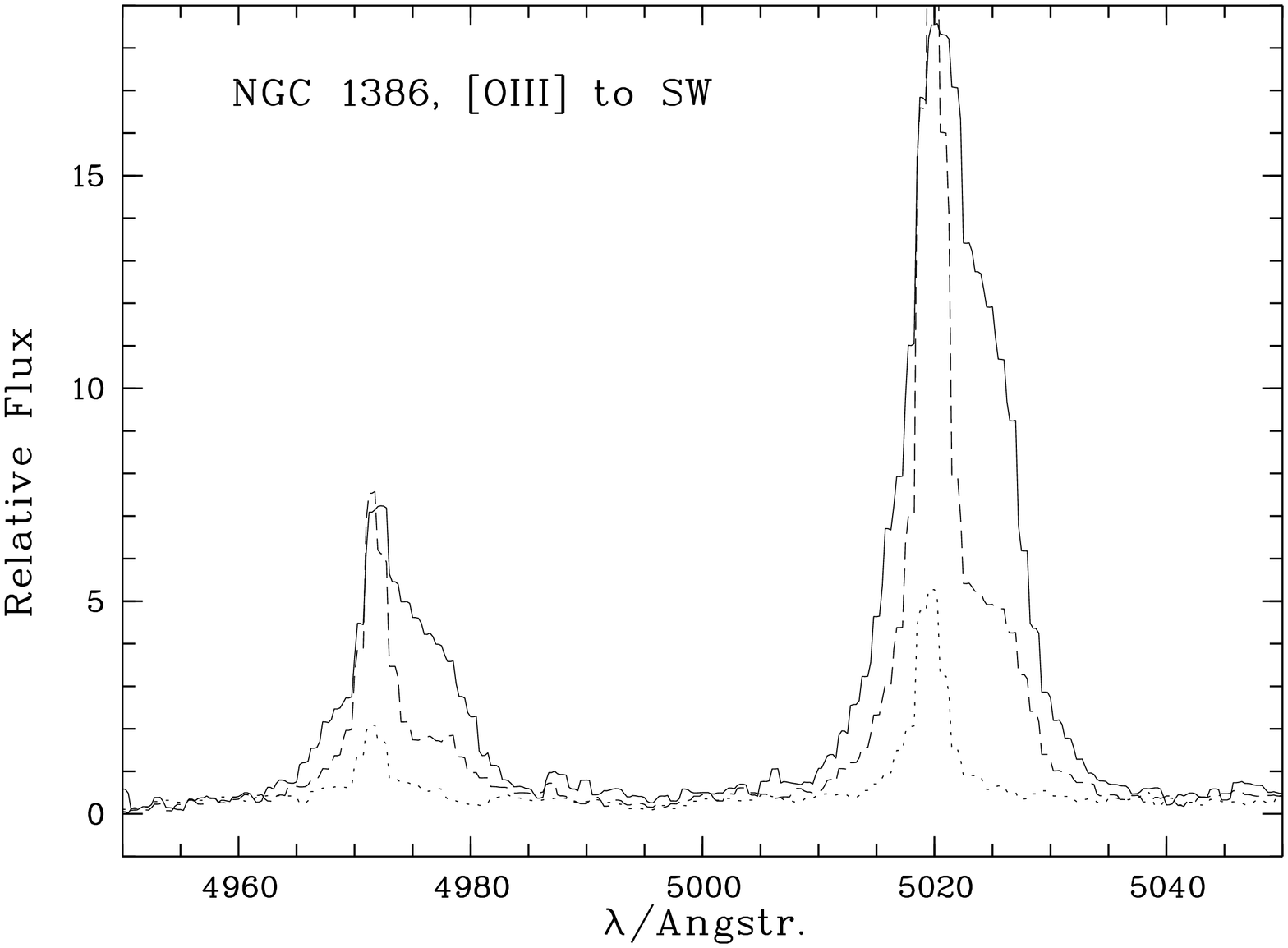}}
\caption{[O{\sc iii}]$\lambda4959$ and $\lambda5007$ line profiles extracted from 1\farcs36 wide 
spectral scans (of spectrum F112). Continuous line: from region $\cal{B}$ (line maximum); 
dashed: 1\farcs36 southwest; dotted: 2\farcs72 southwest.}
\label{o3sw}
\end{figure}

\begin{figure}[ht]
\resizebox{\hsize}{!}{\includegraphics[angle=-90]{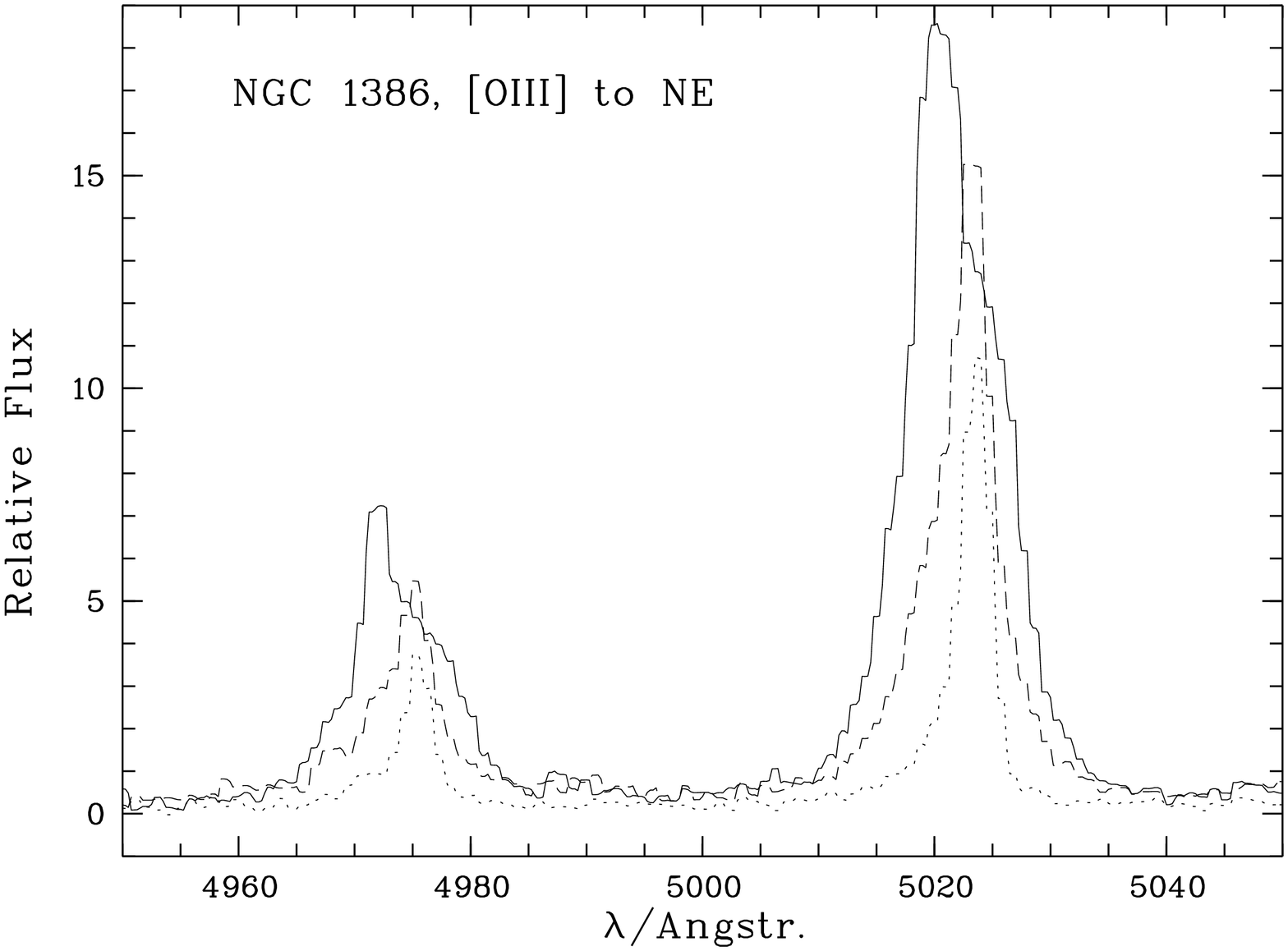}}
\caption{As in Fig.\,\ref{o3sw}, but now towards northeast. Continuous line: from region $\cal{B}$ 
(line maximum); dashed: 1\farcs36 northeast; dotted: 2\farcs72 northeast}
\label{o3ne}
\end{figure}

\begin{figure}[ht]
\resizebox{\hsize}{!}{\includegraphics[angle=-90]{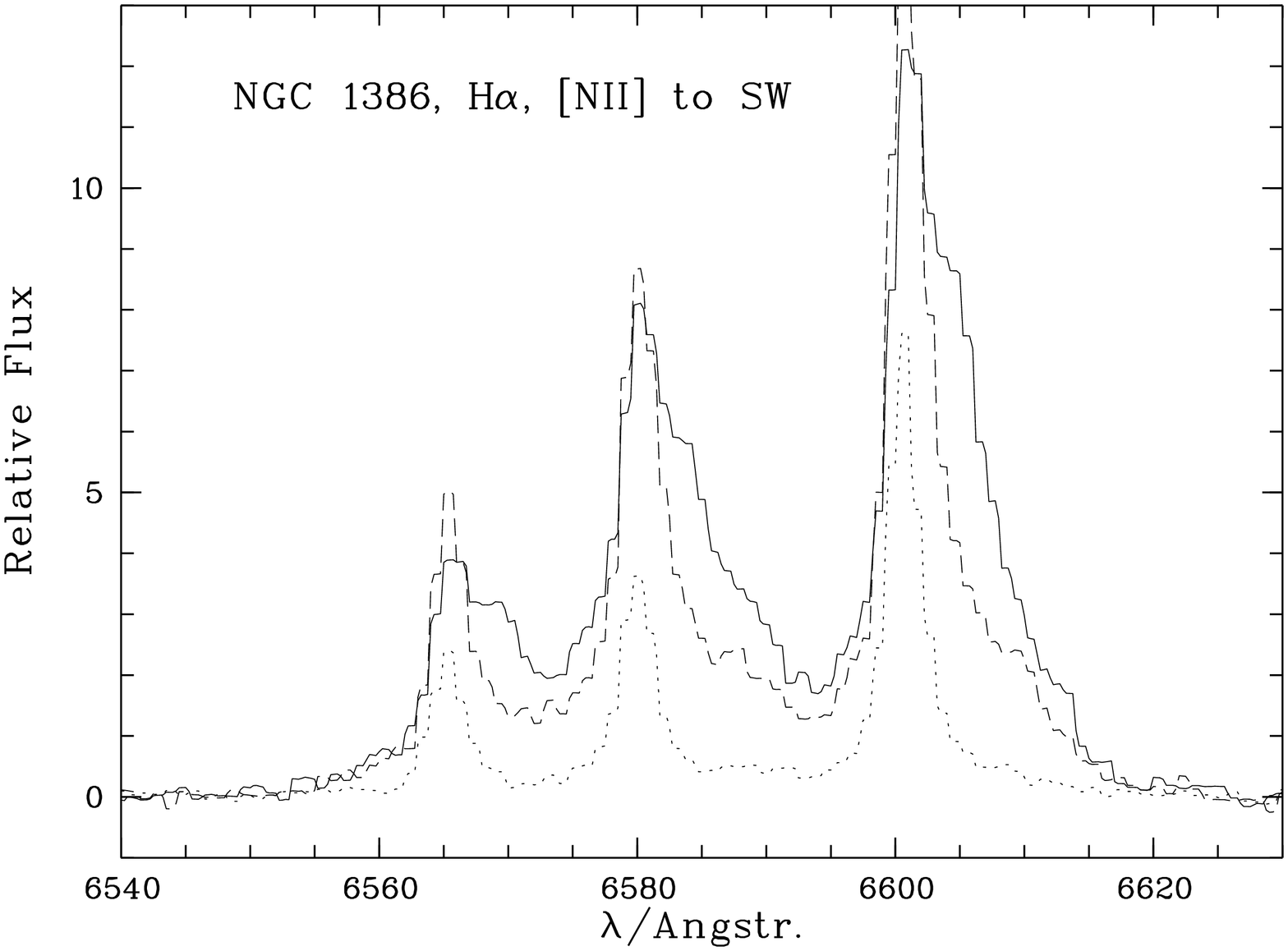}}
\caption{H$\alpha$+[N{\sc ii}]$\lambda\lambda6548, 6583$ line profiles extracted from 1\farcs36 wide 
spectral scans (of spectrum F080). Continuous line: from region $\cal{B}$  (line maximum); dashed: 
1\farcs36 southwest; dotted: 2\farcs72 southwest.}
\label{hasw}
\end{figure}

\begin{figure}[ht]
\resizebox{\hsize}{!}{\includegraphics[angle=-90]{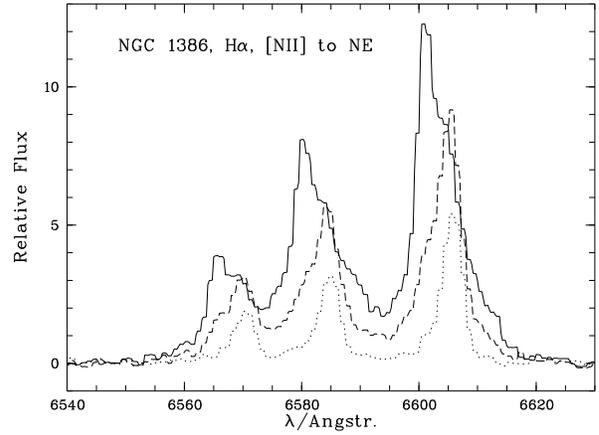}}
\caption{As in Fig.\,\ref{hasw}, but now towards northeast. Continuous line: from region $\cal{B}$  
(line maximum); dashed: 1\farcs36 northeast; dotted: 2\farcs72 northeast.}
\label{hane}
\end{figure}

Figs.\,\ref{o3sw} and \ref{o3ne} show the `blue' line profiles of [O{\sc iii}], and Figs.\,\ref{hasw} 
and \ref{hane} display the `red' line profiles of H$\alpha$ and [N{\sc ii}] from the `central' (max. 
line emission) 1\farcs36 along the slit and the next two 1\farcs36 wide bins towards the southwest 
(p.a.\,(23\degr+ 180\degr)) and towards the northeast (p.a.\,23\degr), respectively. Although the 
blue and red frames were exposed at 1\farcs4 and 2\farcs5 seeing, respectively, the 
profiles of [O{\sc iii}] on one side and H$\alpha$ or [N{\sc ii}] on the other side reveal the same basic 
shape.  The most pronounced difference appears at 1\farcs36 southwest, the sharp shoulders on the 
red side of [O{\sc iii}] are smoothed out in H$\alpha$ and [N{\sc ii}], which we trace back to the seeing effect.

%
%
%
%
%
%
%
%

In the last subsection we concluded that the spectrum along p.a.\,23\degr~arises in an inner 
gaseous disk that is seen near to edge-on and whose line emitting regions are encompassed by 
the 2\arcsec~wide slit. Hence, we expect line profiles resulting from line-of-sight integrations 
through the disk.

All profiles have pronounced narrow peaks (called line cores here) with steep flanks.
The velocities of the narrow-line cores are given in Table\,\ref{c86}. Our spectral resolution
is about half of the apparent width of the line core of disk-component $\cal{B}$. The first 
numerical column in Table\,\ref{c86} gives the velocities of region $\cal{B}$, as measured with 
different lines. As Figs.\,\ref{o3sw} to \ref{hane} or Table\,\ref{c86} show, the radial velocity 
of the core of $\cal{B}$ is close to those at 1\farcs36 southwest and 2\farcs72 southwest, 
while the regions to the northeast are redshifted. Hence, region $\cal{B}$ is on the `blue' 
-- approaching -- side of the rotating disk. In Figs.\,\ref{o3ne}~and \ref{hane}~we see the 
core from $\cal{B}$ and the cores from the northeast line profiles together. The radial velocity 
of the center of the disk (the disk-center velocity) should then be amidst the southwest and 
northeast cores, at $v_{\rm hel}^{\rm c} = 878$\,km/s (mean of six spectral lines). This is in 
agreement with $v_{\rm KC}=877$\,km/s determined from folding the outer rotation curve 
(Sect.\,\ref{glo13}).

%
\begin{table*}
\caption[]{Measured velocities (km/s) of line cores along p.a.\,23\degr~in the center of NGC\,1386. 
Uncertainty 15\,km/s}
\label{c86}
\begin{flushleft}
\begin{tabular}{lccccc}
\hline
Line & At max. (B) & 1\farcs36 southwest &  2\farcs72 southwest & 1\farcs36 northeast & 2\farcs72 northeast  \\
\hline\hline
$[{\rm O}$\,{\sc iii}]$\lambda4959$            & 776 & 757 &  747  &  964 &  983  \\
$[{\rm O}$\,{\sc iii}]$\lambda5007$            & 793 & 762 &  746  &  962 &  984  \\
H$\beta$                                       & 777 & 767 &  751: &  965 &  965: \\
$[{\rm O}$\,{\sc i}]$\lambda6300$              & 849:& 750 &  741  &  947 &  996  \\ 
H$\alpha$                                      & 790 & 777 &  770  &  967 &  993  \\  
$[{\rm N}$\,{\sc ii}]$\lambda6583$             & 789 & 775 &  768  &  980 & 1003  \\
$[{\rm S}$\,{\sc ii}]$\lambda\lambda6716,6731$ & 798 & 776 &  771  &  979 & 1016  \\
\hline
\end{tabular}
\end{flushleft}
\end{table*}
%

Schulz et al. (1995) (their Fig.\,4) showed that a rotating edge-on disk with an emissivity 
law falling with radius will produce asymmetric profiles, with the more extended wing pointing 
towards smaller absolute values of rotation velocity. However, the wings should not pass the 
disk-center velocity unless there are strong random motions and/or radiation `from the other 
side' is scattered in via observational seeing (and guiding inaccuracies), as is shown by an 
example in Sect.\,7.2 of Schulz et al. (1995).

The asymmetric line profiles of component $\cal{B}$ have the extended wing on the side expected 
for a rotating disk, but their shoulder extends to longer wavelengths than theoretically predicted 
unless there is strong spatial smearing or an inner high-velocity branch along the same l.o.s. 
(e.g. central Keplerian rotation). The shoulders even extend beyond the northeast-cores (Figs. 
\ref{o3ne} and \ref{hane}). Note that in velocity units the red flank of the core 1\farcs36 
northeast lies at about the upper limit of the rotation curve in Fig.\,\ref{v13} ($\sim 1050$\,km/s) 
and the blue flank of the core of component $\cal{B}$ (or the core of the component 
1\farcs36 southwest) lies at the lower limit of the rotation curve ($\sim 700$\,km/s). 
Consequently, the outer red shoulder of profile component $\cal B$ cannot be explained by 
light from regions of the outer rotation curve, while the sharp transitions to the inner part 
of the red shoulder (particularly in H$\alpha$ and [N{\sc ii}] of $\cal B$) and in [O{\sc iii}] 1\farcs36 
southwest suggest that this is emission from the northeast side of the disk. This picture implies 
that there is nearly no emission seen from the nucleus of the disk at the disk-center velocity
(in agreement with the location of KC within the dust lane; see Sect.\,\ref{glo13}) as such 
emission would have smoothed out the transition from the southwest to the northeast component. 

Outer wings are found on the blue and red side of component $\cal{B}$ and extend up to 900\,km/s 
symmetrically to the disk-center velocity in [O{\sc iii}], but not symmetrically to the core velocity 
of $\cal{B}$. Wings can also be traced to a slightly lesser extent from the components at 
1\farcs36 and 2\farcs72 distance from $\cal{B}$. In [N{\sc ii}] outer wings appear to extend to 
$\sim 700$\,km/s. 

A possible scattering origin of such wings will be discussed in Sect.\,\ref{sca}. Since the
kinematical center is obscured by the dust lane, an explanation by fast Keplerian motion
close to the disk center is unlikely.

Spectrum F023 taken along p.a. 90\degr~(along which FWM2000 detected two emission-line plumes) 
shows emission over $\pm2\arcsec$~ with a slightly declining velocity gradient from east to west 
($\sim 5$\,km/s/\arcsec). The near-north-south orientation of the disk explains why a notable 
east-west gradient in the l.o.s. projections of rotational motions is unexpected. Its near-edge-on 
orientation makes the detection of minor-axis outflow difficult. The gradient could be due to 
outflow along the plumes if the disk is slightly tilted from edge-on. The extended red wing in 
H$\alpha$, [N{\sc ii}] and [S{\sc ii}] (seen in the F080 spectrum of component B (continuous line) in
Figs.\,\ref{hasw} and \ref{hane}) is confirmed by F023. 

\subsubsection{Line ratios}
\label{lin13}
Albeit of the primarily kinematic nature of the present study, a few relative intensities of the
major diagnostic lines are to be noticed in order to identify the nature of the emission-line 
region in question. Weaver et al. (1991) noticed a wealth of varying emission-line ratios from 
the typical Sy-2 high-excitation ratios to extranuclear LINER-like ratios. Even in the few 
strips that our spectra cover, drastic line-ratio changes are seen. Along the p.a. (23\degr, 
23\degr+180\degr)  slit, [N{\sc ii}]$\lambda6583$/H$\alpha$, [S{\sc ii}]($\lambda6716$+$\lambda6731$)/H$\alpha$ 
and [O{\sc i}]$\lambda6300$/H$\alpha$ increase from 1.5, 0.68 and 0.13 at the brightness maximum 
($\cal B$) to 2.5, 1.10, 0.23, respectively, at 4\arcsec~southwest. To the northeast there 
is only a moderate (10\%) increase towards 3\arcsec~northeast in [N{\sc ii}] and [S{\sc ii}], only 
[O{\sc i}] rises to 0.2. 

[O{\sc iii}]/H$\beta$ is 13 at $\cal B$, but rises to 21 at 1\farcs36 southwest, and 17 at 1\farcs36 
northeast. [S{\sc ii}]$\lambda6716$/[S{\sc ii}]$\lambda6731$ is 1.0 at $\cal B$ (indicating $n_e = 7\,10^2$ 
cm$^{-3}$) and 1\farcs36 southwest, but rises to 1.2 for the next 3\arcsec~on either side 
(indicating an average $n_e = 2.5\,10^2$ cm$^{-3}$).

A faint spectrum along p.a. 90\degr~(F026), with the slit set 2\arcsec~north of $\cal B$, shows 
essentially constant values $\sim 1$ and $\sim 0.8$ ($n_e \approx 10^3$ cm$^{-3}$) for 
[N{\sc ii}]$\lambda6583$/H$\alpha$ and [S{\sc ii}]$\lambda6716$/[S{\sc ii}]$\lambda6731$ over 
about 10\arcsec, respectively.

\subsection{The LINER NGC\,1052} 
%
%
\begin{figure}[ht]
\resizebox{\hsize}{!}{\includegraphics[angle=-90]{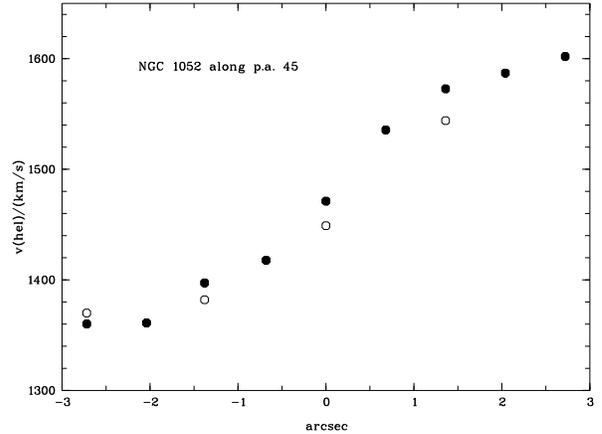}}
\caption{Heliocentric velocities/(km/s) along p.a.\,45\degr in the center of NGC\,1052 (northeast 
to the left). The filled octagons show the mean positions of the line peaks of H$\alpha$ and 
[N{\sc ii}]$\lambda6583$ as measured in the raw spectrum. The open octagons display the mean values 
of single-component Lorentzian fits given in Table \ref{vs52}.}
\label{v52}
\end{figure}
%
\begin{figure}[ht]
\resizebox{\hsize}{!}{\includegraphics[angle=-90]{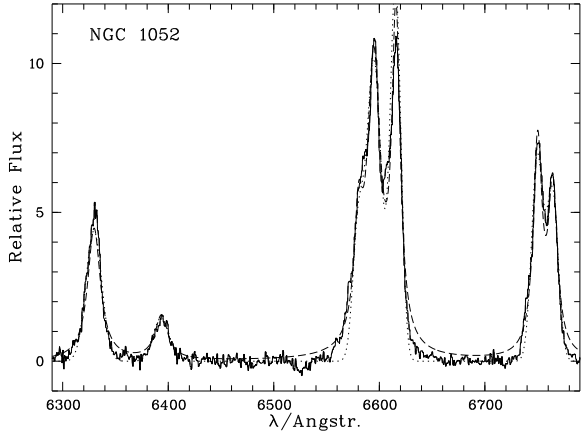}}
\caption{Red spectrum (including [O{\sc i}]$\lambda\lambda6300,6363$, 
H$\alpha$+[N{\sc ii}]$\lambda\lambda6548,6583$, and [S{\sc ii}]$\lambda\lambda6716, 6731$) from the brightness 
maximum ($1\farcs36 \times 2\arcsec$) of NGC\,1052 with the slit aligned along p.a.45\degr. Fits 
with single Lorentzians (dashed) or Gaussians (dotted) are insufficient. Positions and line 
widths from the fits with Lorentzians are given in Table \ref{vs52} (column headed by 0).}
\label{sin52}
\end{figure}
%

%
\begin{figure}[ht]
\resizebox{\hsize}{!}{\includegraphics[angle=-90]{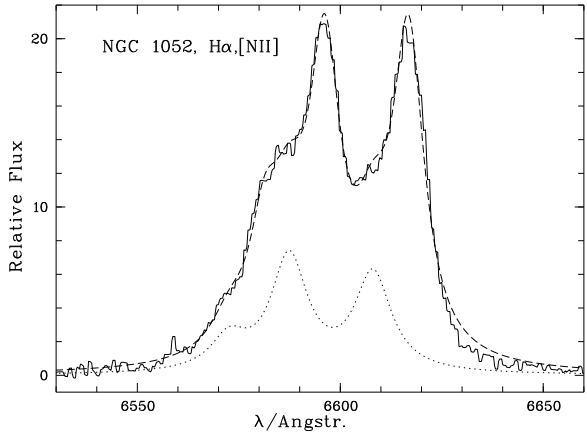}}
\caption{Two narrow-line systems $\cal{A}$ and $\cal{B}$, each characterized by three Lorentzian 
functions with fixed line width and position in velocity space (the intensity ratio of $\lambda6583$ 
to $\lambda6548$ is fixed to 3 to 1), are sufficient to fit the observed 
H$\alpha$+[N{\sc ii}]$\lambda\lambda6548, 6583$~blend from the region with maximum emission (continuous 
line shown above) and till 2\arcsec~northeast. H$\alpha$+[N{\sc ii}] from system $\cal{B}$ (dotted) 
and the `fitting' sum  $\cal{A}+\cal{B}$ (dashed) are shown. While $\cal{B}$ is blueshifted and 
indicates outflow, $\cal{A}$ is consistent to arise in a rotating disk.}
\label{hal52}
\end{figure}
%
%
\begin{figure}[ht]
\resizebox{\hsize}{!}{\includegraphics[angle=-90]{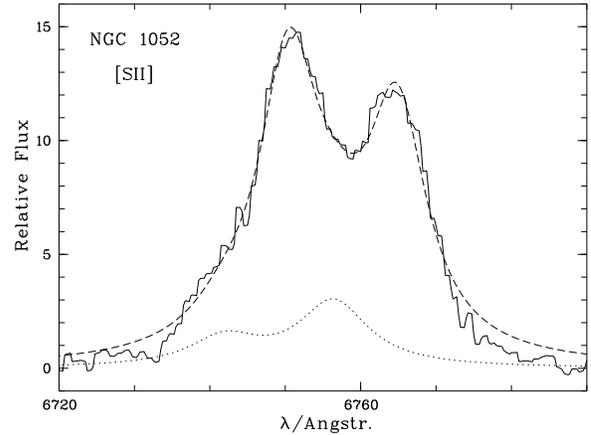}}
\caption{As in Fig.\,\ref{hal52}, but now for [S{\sc ii}]$\lambda\lambda6716, 6731$. Component $\cal{B}$ 
is additionally constrained to be in the high-density limit ($n_e \gg 10^3$ cm$^{-3}$). }
\label{sul52}
\end{figure}
%
%
\begin{figure}[ht]
\resizebox{\hsize}{!}{\includegraphics[angle=-90]{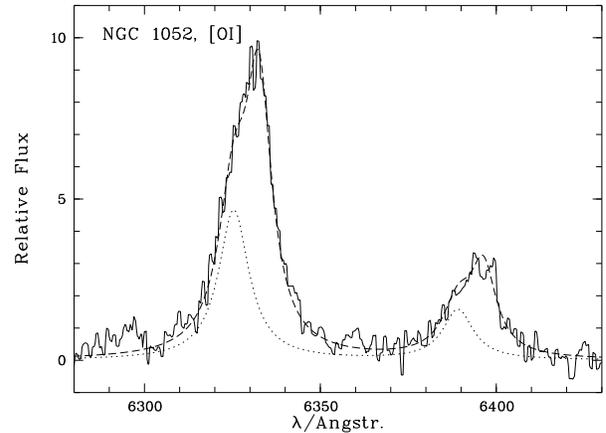}}
\caption{As in Fig.\,\ref{hal52}, but now for [O{\sc i}]$\lambda\lambda6300, 6363$.}
\label{ox52}
\end{figure}
%
%
\begin{figure}[ht]
\resizebox{\hsize}{!}{\includegraphics[angle=-90]{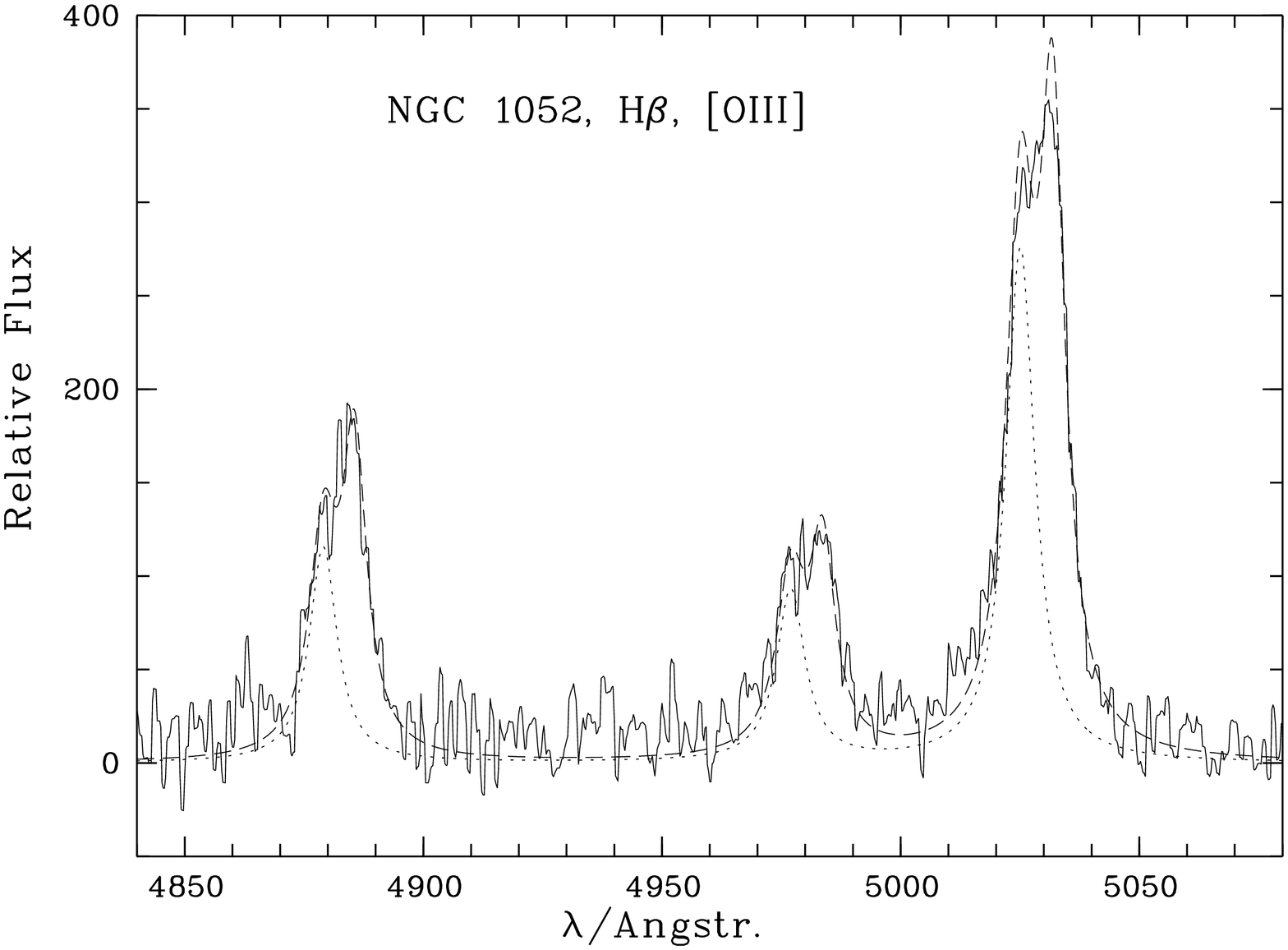}}
\caption{As in Fig.\,\ref{hal52}, but now for H$\beta$ and [O{\sc iii}]$\lambda\lambda4959, 5007$.}
\label{o3b52}
\end{figure}
%
%
%
%
\subsubsection{Fitting line profiles with single components}
\label{fitt}
Davies \& Illingworth (1986) gave p.a. 49\degr~as the approximate position angle of the major axis 
of the nuclear gaseous disk in NGC\,1052. For deprojecting the observed velocities we assume that 
spectra F078 and F107 (slit along p.a.\,$45^{\circ}$) were taken along the major axis. 

Fig.\,\ref{v52} shows two major-axis velocity curves; one shows the positions of the {\em peaks} 
of H$\alpha$ and [N{\sc ii}]$\lambda6583$ in the raw spectra from 0\farcs68 row extractions. The other 
curve corresponds to template-corrected spectra binned to 1\farcs36 wide spectral extractions and
displays the mean values of {\em single-component Lorentzian fits} (see Table\,\ref{vs52}) of the 
lines H$\alpha$, [N{\sc ii}]$\lambda\lambda6548,6583$, and [S{\sc ii}]$\lambda\lambda6716,6731$ from the disk 
(see Appendix B regarding the fitting procedures with Gaussians and Lorentzians as basic functions). 
Beyond the steep central gradient the curves flatten out. They are not symmetric relative to the 
zero point, which is located at the brightness maximum.

Davies \& Illingworth (1986) identified a nuclear ionized disk that emits most of the emission-line 
spectrum. For subsequent mass estimates an inclination of the disk $i=60\degr$ is assumed as derived 
in the following subsection (note that Bertola et al. (1993) obtained $i=53\degr$ from the data of
Davies \& Illingworth; this value would increase the mass estimates given below by 18\%).

The central velocity gradient (between maximum and two adjacent pixels) of 85\,km/s/\arcsec~  
translates formally into 59.2\,km/s/(50pc), yielding a Keplerian mass of $4.1\,10^7$\,M$_{\odot}$ 
inside a radius of 50\,pc, which corresponds to the inner, just resolved core with 1\farcs2 diameter.

The processed data (Table \ref{vs52}) have 1\farcs36 resolution and lead to a gradient of 110\,km/s
per 2\arcsec~radius (measured at a velocity of 1478\,km/s), which translates into $6.3\,10^8$\,M$_{\odot}$ 
inside a radius of 166 pc.

Albeit crude, these mass estimates imply that NGC\,1052 has a core of relatively small mass. In order 
to judge the reliability of these numbers a closer look on the fits is needed. 

The data given in Table \ref{vs52} were obtained by the Lorentzian fits of [O{\sc i}]$\lambda\lambda 6300, 
6363$, H$\alpha$, [N{\sc ii}]$\lambda\lambda 6548, 6583$, and [S{\sc ii}]$\lambda\lambda 6716, 6731$ shown in 
Fig.\,\ref{sin52}.  Although fitting the major part of the profiles these fits are unsatisfactory 
in several respects. 

Firstly, measured outer line wings appear too weak or too strong as compared to Lorentzian or Gaussian 
fits, respectively. In itself, this might not be taken as serious because it could simply reflect 
the deficiency of these functions in comparison to the typical core-wing structure of narrow-line 
profiles.

Secondly, the final summed fit lies systematically below the bridges connecting [N{\sc ii}]$\lambda6548$ 
and H$\alpha$ or [N{\sc ii}]$\lambda6583$ and H$\alpha$, and connecting the [S{\sc ii}] lines. It also 
lies below the blue wing of [N{\sc ii}]$\lambda 6548$. This holds irrespective of the type of profile function.

Thirdly, although no conspicuous asymmetry is seen in [O{\sc i}], in this case the fit function turns out 
to be blueshifted by $\sim 100$\,km/s and appears 80\% wider with respect to other forbidden lines. 
Also, sometimes H$\alpha$ requires considerably larger widths than [N{\sc ii}] although its position is 
not significantly shifted (Table \ref{vs52}).

These findings indicate that the line profiles are complex and that a one component model is not 
sufficient for an appropriate simulation. In the next subsection we therefore decompose the line
profiles in a more detailed manner.

%
\begin{table*}
\caption[]{This table gives heliocentric velocities and (after the $|$) values of $w$=FWHM (in km/s) 
derived by single-component Lorentzian fits in template-corrected 1\farcs36 wide spectral extractions along 
p.a.\,45\degr~of NGC\,1052. For [O{\sc i}] and [N{\sc ii}] the doublet ratios 3:1 and the requirement of same 
widths (as in [S{\sc ii}]) additionally constrain the fits. The column labeled `0' refers to the 
1\farcs36 wide spectral scan at maximum emission. Last row: mean values of the lines H$\alpha$, 
[N{\sc ii}]$\lambda\lambda6548,6583$, and [S{\sc ii}]$\lambda\lambda6716,6731$ (open octagons in 
Fig.\,\ref{v52}). }
\label{vs52}
\begin{flushleft}
\begin{tabular}{lccccccccc}
\hline
Line & 5\farcs44\,NE & 4\farcs08\,NE &  2\farcs72\,NE) & 1\farcs36\,NE & 
0 & 1\farcs36\,SW & 2\farcs72\,SW & 4\farcs08\,SW & 5\farcs44\,SW \\
\hline\hline
H$\alpha$                 & $1330|161$ & $1369|137$ &  $1369|226$  &  $1376|492$ &  $1430|706$
& $1531|545$ & $1594|270$ & $1615|196$ & $1614|172$  \\
$[{\rm N}$\,{\sc ii}]     & $1363|168$ & $1381|128$ & $1366|205$ &  $1380|316$  &  $1446|435$ 
& $1552|383$ & $1607|299$ & $1631|229$ & $1629|201$\\
$[{\rm S}$\,{\sc ii}]     & $1342|178$ & $1374|156$ & $1374|229$  &  $1391|365$  & $1471|454$
& $1550|393$ & $1605|264$ & $1645|208$ & $1621|200$  \\
$[{\rm O}$\,{\sc i}]$\lambda6300$ & --- & $1361|360$ & $1325|401$   & $1330|647$ & $1373|711$
& $1469|709$ & $1622|331$ &   ---      & ---         \\
\hline
mean & $1345|169$ & $1375|140$  &  $1370|220$  &  $1382|391$  & $1449|532$
& $1544|440$ & $1602|278$ & $1630|211$ & $1621|191$  \\  
\hline
\end{tabular}
\end{flushleft}
\end{table*}
%
%
\subsubsection{Decomposition of the line profiles}
In view of the problems encountered with single-component fits, a significant improvement may 
be obtained by a mixture of two line systems, conveniently denoted as $\cal A$ and $\cal B$, 
with system $\cal B$ blueshifted with respect to $\cal A$.

In Fig.\,\ref{hal52} a fit to H$\alpha$ + [N{\sc ii}] from the `central' (center defined by maximum 
emission) $2\arcsec \times 0\farcs68$ (slit width $\times$ width of CCD extraction row) is 
shown along p.a. 45$^{\circ}$. The line blend can indeed be successfully decomposed into 
two narrow-line H$\alpha$ + [N{\sc ii}] systems. However, such a fit of a smeared out blend of 
lines can only be found by the software if guided by strong constraints, which we are 
now going to justify.

In Table\,\ref{ab52} final results for the FWHMs, velocities and  line ratios of systems 
$\cal A$  and $\cal B$ are given. The data are restricted to the spatial regime between 
2\arcsec~northeast and 0\farcs68 southwest for which component $\cal B$ can be unambiguously  
recognized. 
%
%
%
\begin{table*}
\caption[]{Line widths $w$=FWHM, heliocentric velocities and line ratios measured for the 
rotational component $\cal{A}$ and the outflow component $\cal{B}$ (separated by a $\vert$) 
along p.a.\,45\degr~from 0\farcs68 southwest to 2\farcs04 northeast in NGC\,1052.}
\label{ab52}
\begin{flushleft}
\begin{tabular}{llllll}
     & 2\farcs04 NE & 1\farcs36 NE &  0\farcs68 NE & 0  & 0\farcs68 SW  \\
\hline\hline
$w({\cal A}) \vert w({\cal B})$ & $268 | 208$ & $296 | 366$ & $378 | 435$ & $420 | 498$ & $340 | 565$ \\
\hline\hline
Line     &        &       &       &       &       \\
\hline\hline
H$\beta$, [O{\sc iii}]   & $1354 | 931$ & $1407 | 1059$ &  $1468 | 1064$ &  $1496 | 1081$ &  $1629 | 1342$  \\
H$\alpha$, [N{\sc ii}]   & $1371 | 998$ & $1414 | 1043$ &  $1449 | 1066$ &  $1505 | 1096$ &  $1567 | 1249$  \\
$[{\rm S}$\,{\sc ii}]    & $1369 | $ ---& $1414 | 1109$ &  $1448 | 1066$ &  $1498 | 1154$ &  $1559 | 1353$  \\
$[{\rm O}$\,{\sc i}]     & $1355 | 992$ & $1397 | 1016$ &  $1484 | 1141$ &  $1506 | 1165$ &  $1575 | 1256$  \\
\hline 
mean & $1364 | 980$ & $1409 | 1054$ & $1460 | 1081$ & $1502 | 1116$ & $1579 | 1290 $\\
mean$-1502$ & $-138|-522 $ & $-93|-448$  & $-42|-421$ & $ 0|-386 $ & $77| -212$ \\
\hline\hline 
$I(\lambda5007) / I$(H$\beta$)        & $1.97 | 1.94$ & $1.74 | 2.69$ &  $2.03 | 2.43$  &  $1.71 | 2.82$  & $2.03 | 2.15$  \\  
$I(\lambda6583) / I$(H$\alpha$)  & $1.16 | 1.37$ & $1.12 | 1.32$ &  $1.05 | 1.09$  &  $1.09 | 0.86$ & $1.23 | 0.89$ \\
$I$([S{\sc ii}]$_{\rm tot} / I$(H$\alpha$)   & $1.29 | $ --- & $1.13 | 0.90 $ & $1.18 | 0.84 $ & $1.36 | 0.62$ & $1.58 | 0.91$ \\
$I(\lambda6300) / I$(H$\alpha$) & $0.48 | 1.79$ & $0.58 | 0.57$ & $0.48 | 0.87$ & $0.45 | 0.66$ & $0.39 | 0.59$ \\ 
\hline\hline
$I(\lambda6716)/I(\lambda6731)$($\cal{A}$) & 1.40 & 1.15 & 1.04 & 1.17 &  \\
\hline
\end{tabular}
\end{flushleft}
\end{table*}
%
%
%
%

The H$\alpha$+[N{\sc ii}] blend has been fit by assuming that all three lines in each system are 
constrained to have the same radial velocity and line width $w$ (=FWHM; see Appendix B) while 
the intensities of H$\alpha$ and [N{\sc ii}] are left free (the 3:1-[N{\sc ii}] ratio is fixed; 
see Appendix B). Each of the two systems is fixed by four parameters (same position and same width in 
velocity space, height of H$\alpha$ and height of [N{\sc ii}]$\lambda6548$). Thus, only eight 
parameters are fit for six line profiles which would  be determined by eighteen parameters 
in total if each single profile were free of constraints.
 

A slight shoulder on the blue wing of [S{\sc ii}]$\lambda 6716$ requires the presence of the 
blueshifted component in [S{\sc ii}] as well. An example for a fit of the [S{\sc ii}] system employing 
analogous precepts as for H$\alpha$ + [N{\sc ii}] (adopting $w(\cal A)$ and $w(\cal B)$~from the 
H$\alpha$ + [N{\sc ii}] fit of each region) is shown in Fig.\,\ref{sul52}. Five out of twelve
parameters were fit (two velocities, for systems $\cal{A}$ and $\cal{B}$, two amplitudes
for system $\cal{A}$ and one amplitude for system $\cal{B}$, where the [S{\sc ii}] intensity 
ratio was fixed to $I(6731)/I(6716)=2.3$, which is the saturation limit for electron densities 
$n_e \gg 10^3$ cm$^{-3}$). 

A large electron density of $\cal{B}$ has been chosen to increase the contribution of 
[O{\sc i}]$\lambda\lambda6300,6363$ relative to [S{\sc ii}]$\lambda\lambda6716,6731$. The critical 
density of the upper [O{\sc i}] levels, $1.7\,10^6$ cm$^{-3}$, is three orders of magnitude above
that for [S{\sc ii}]. A high electron density is indicated by the single-component fits which 
yield large relative strengths of [O{\sc i}] in the diagnostic diagrams of Veilleux \& Osterbrock 
(1987; hereafter VO) and larger line widths than those of [S{\sc ii}] and [N{\sc ii}] (Table \ref{vs52}). 
The trend that [O{\sc i}] can be boosted mainly by density is shown in Komossa \& Schulz (1997; 
their Fig.\,4, upper panel). Even though a distinct blue peak or a clear shoulder is missing in 
[O{\sc i}]$\lambda6300$ (Fig.\,\ref{ox52}), the two-component fit with Lorentzian functions is 
satisfactory and explains the greater width of the [O{\sc i}] profiles by a relatively strong $\cal{B}$ 
contribution.

Due to the likewise high critical density of $7\,10^5$\,cm$^{-3}$ of the $^1{\rm D}_2$ upper term of 
[O{\sc iii}]$\lambda\lambda 4959, 5007$, these lines may be density-boosted in high-ionization zones.
Consequently component $\cal B$ might be conspicuous in these lines as well which is indeed 
verified by a significant blue shoulder of the profiles (Fig.\,\ref{o3b52}). The [O{\sc i}] and 
[O{\sc iii}] spectra were consequently fit in the same way as the [S{\sc ii}] lines; making use 
of 3:1-line ratios in each of the oxygen doublets, however, four free parameters out of a total 
of twelve were sufficient, one amplitude and one velocity for each of the two systems, $\cal{A}$ 
and $\cal{B}$, respectively. 

The row `mean' in Table\,\ref{ab52} gives the average velocities over the four line groups
(H$\beta$+[O{\sc iii}], [O{\sc i}], [S{\sc ii}], H$\alpha$+[N{\sc ii}], the latter is weighted 
twice) for velocity systems $\cal{A}$ and $\cal{B}$. The line velocities for system $\cal{A}$ given in 
Table\,\ref{ab52} are consistent with arising in a rotating circumnuclear disk (as suggested 
by Davies \& Illingworth 1986). The differences relative to 1502\,km/s (next row) exhibit a nearly 
linear gradient to the northeast of 57\,km/s per 0\farcs68 step (=56.37 pc) or 51 km/s at a formal 
radius of 50 pc from the kinematical center (which appears to be slightly shifted from the location 
of maximum emission). This would lead to a Keplerian mass estimate of $3\,10^7$\,M$_{\odot}$ inside 
50 pc if the disk were edge-on, or $4\,10^7$\,M$_{\odot}$ for $i = 60\degr$ as given below. The latter 
value is in agreement with the single component fits given in Sect.\,\ref{fitt} because the blue 
component $\cal{B}$ hardly effects the overall fits (except for [O{\sc i}] which was not considered in 
the mean values of Table\,\ref{vs52}).

Two slightly poorer spectra along p.a.\,90\degr (F020 and F056) were decomposed in a similar way
(with outflow fixed by 400\,km/s). From these a component--$\cal{A}$ gradient of 15\,km/s per 
0\farcs68 was found. With p.a.\,(major axis) = 45\degr~we can adjust the velocity deprojection 
factor of Eq.\,\ref{rot} so that along both slit directions the rotational velocities in the 
disk agree. This procedure leads to an inclination angle $i = 60\degr \pm 5\degr$.  

The blueshifted system $\cal{B}$ is most likely explained by the presence of outflowing gas. At 
various p.a. blueshifted structure in the wings of [O{\sc iii}] were earlier described by Davies 
\& Illingworth (1986).  Although they had not attempted a quantitative numerical decomposition
of their line profiles, their insightful discussion led to the detection of widespread
outflow on the eastern side of the nucleus. 

The spatial covering of the present data is insufficient for clues about specific outflow 
geometries of component $\cal{B}$. Also, prior to deeper interpretation the degree of uniqueness 
of the fit-procedure has to be discussed. To this end, we had also attempted fits with other
constraints, e.g. fixing the blueshift of $\cal{B}$ (at 400\,km/s relative to $\cal{A}$) rather than 
fixing the line width. The kinematics of the rotational component $\cal{A}$ was hardly altered 
suggesting that common  velocities and linewidths in lines from different ions are a robust 
feature for a bulk component. 

Despite general agreement, to the northeast Table\,\ref{ab52} shows differences in the 
outflow velocities of up to $\sim70$\,km/s. Such differences are within the accuracy of the 
fitting procedure that employs a simple basic function. Note that most differences are 
significantly below the component widths and the spectral resolution. The worst discrepancies 
are at 0\farcs68 southwest, but here the outflow component starts to vanish. 

Row `mean--1502' in Table\,\ref{ab52} gives the l.o.s. components of the outflow velocity 
relative to the emissivity maximum, identified at 1502\,km/s. On the northeastern side 
it is blueshifted by $\sim (450\pm70)$\,km/s; at 0\farcs68 southwest less consistent fits 
yield a blueshift of $\sim200$\,km/s. 

With a disk inclination angle of 60\degr~the northeast outflow velocity would be $\sim 900$\,km/s 
relative to the emissivity maximum if it were directed along the minor axis. However, a detailed 
deprojection has to await more complete data. 
\subsection{The Seyfert Mrk\,1210}
\subsubsection{The WR-feature} 
The three spectra extracted from F116 (Table 1) with $2\arcsec$~wide bins from the central 
$6 \arcsec$~(diameter) exhibit HeII$\lambda 4686$ with an observed intensity of 25\% of H$\beta$. 

Concentrated in the central 2\arcsec~we detect a Wolf-Rayet feature around $\lambda 4686$ 
(for Mrk\,1210 first shown in Storchi-Bergmann et al. 1998).

Considering individual features, we note that [Fe{\sc iii}]$\lambda 4658.10$ is clearly present, 
[Ar{\sc iv}]$\lambda 4740.34$ is present outside the innermost $2\arcsec$, while [Ar{\sc iv}]$\lambda 
4711.34$ and [Ne{\sc iv}]$\lambda 4724.15$ are possibly present, but cannot be clearly discerned 
at our signal-to-noise level.

The WR-feature is an unambiguous sign for the presence of a starburst which appears to arise 
cospatial with the NLR or in a region roughly seen in projection to the inner kiloparsec in 
which the bright part of the AGN NLR is situated.
\subsubsection{Analysis based on single narrow-line component}
\label{fou}
\begin{figure}[ht]
\resizebox{\hsize}{!}{\includegraphics[angle=-90]{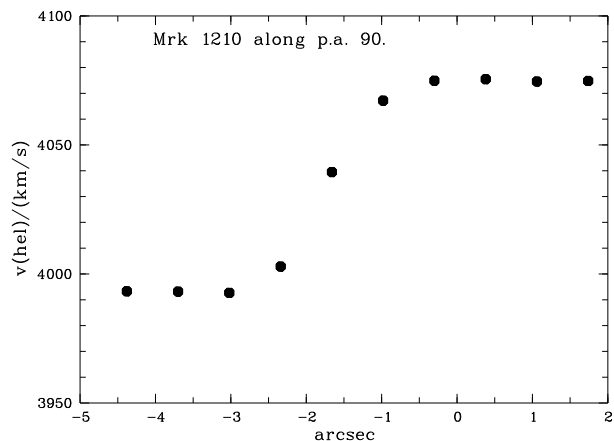}}
\caption{Velocity curve (heliocentric) of Mrk\,1210 based on the mean of the peak positions of 
H$\alpha$ and [N{\sc ii}]$\lambda6583$ in the raw spectrum F066 along p.a. 90\degr (east to the left). 
The brightness maximum is at the zero position. The curve is smoothed by spatial smearing 
($>2\arcsec$).}
\label{vmk}
\end{figure}

Firstly, peak positions of the strongest lines were measured. The derived nuclear velocity 
curve along p.a. 90\degr~ yields a small velocity gradient (Fig. \ref{vmk}). The gradient is 
flattened because of seeing and guiding inaccuracies of $\sim 2\arcsec$. A deconvolution is 
not warranted because of the lack of an appropriate point-spread function.

Locating the kinematical center KC midway within the slope yields a measured difference to the 
brightness maximum BM of 1\farcs5. In the weaker blue spectrum (F116) the slope plus three
extraction rows westwards can be measured in [O{\sc iii}]. Here the difference between KC and the
[O{\sc iii}] BM amounts to two rows or 1\farcs3. Due to the spatial smearing the real difference
is likely $\sim 1\arcsec$ or slightly less.

From velocities measured in the weak spectra (F085, F087), which were obtained offset from the 
brightness center by 2\arcsec~(slit along p.a. 90\degr) we estimate a p.a. of the kinematical 
major axis of $110 \degr \pm 20\degr$. The uncertainty of the p.a. is due to the faintness of
the extranuclear lines. 

From the observed slight velocity variations one cannot safely distinguish between bipolar 
outflow and rotation. We tend to regard it as a reflex from the rotating nuclear kpc-spiral 
disk seen in a snapshot-HST-image (Malkan et al. 1998). However, neither from the kinematical 
data nor from the printed HST image a meaningful value of inclination can be derived, so that 
no reliable mass estimate can be given. 
%
%
%
\begin{table*}
\label{mks}
\caption[]{Heliocentric velocities and line widths (FWHM) (separated by $\vert$~) in km/s 
along p.a. \,90\degr~through the nucleus of Mrk\,1210 as derived from single-component Lorentzian fits. 
The zero position is that of the interpolated brightness maximum (symmetry center of brightness).}
\label{mks}
\begin{flushleft}
\begin{tabular}{llllllll}
Position    & H$\beta$ &[O{\sc iii}]&[O{\sc i}]& H$\alpha_{\rm broad}$ & H$\alpha_{\rm narrow}$ & [N{\sc ii}] & [S{\sc ii}]  \\
\hline\hline
3\farcs02 E &          &   &  $3992 | 809$ &  $3999 | 2268:$ &  $3999 | 116$ & $3998 | 137$ & $3999 | 148$  \\
2\farcs34 E & $4008 | 329$ & $4001 | 471$ &   $4076 | 715$  & $4004 | 1893$  &  $4004 | 188$  & $4019 | 213$ & $4017 | 177$  \\
1\farcs66 E   & $4021 | 450$ & $4034 | 507$ & $4104 | 605$ & $4048 | 1759$ &$4048 |313$   & $4052 | 296$ 
& $4053 | 256$ \\
\hline 
0\farcs98 E & $4056 | 557$ & $4054 | 495$ &  $4127 | 617$  &  $4076 | 1733$  & $4076 | 331$ & $4081 | 287$ &
 $4072 | 274$  \\
0\farcs30 E& $4066 | 595$ & $4063 | 494$ & $4133 | 607$& $4088 | 1707$ & $4088 | 351$ & $4092 | 319$ &
 $4081 | 276$ \\
0\farcs38 W  & $4043 | 486$ & $4051 | 572$ & $4135 | 587$ & $4089 | 1703$ & $4089 | 392$ & $4095 | 328$ & 
$4083 | 271$ \\
1\farcs06 W   & $4055 | 474$ & $4058 | 505$ & $4126 | 631$ & $4084 | 1684$& $4084 | 396$ & $4091 | 319 $& 
$4084 | 275$\\
\hline
1\farcs74 W   & $4017 | 421$ & $4042 | 478$ &  $4097 | 640$  &  $4078 | 2429$  & $4078 | 267$ & $4074 | 212$ &
$ 4088 | 227$  \\  
2\farcs42 W   &    &      &      &  $4092 | 5129:$ & $4092 | 137$ & $4076 | 144$ & $4094 | 166$  \\
\hline
\end{tabular}
\end{flushleft}
\end{table*}

Leaving aside the global kinematics, various trial fits of the strongest line profiles (observed to the 
west of the location of the velocity gradient) were undertaken which led to reasonably successful
single-component fits. For justification of the results given in Table\,\ref{mks} and, in particular for an
alternative decomposition of the profiles described in the next subsection, some details of the procedures
employed are to be explained. 
 
Firstly, it was attempted to fit the apparent three-line blends H$\alpha$+[N{\sc ii}] by either a sum of three 
Gaussians or three Lorentzian functions only constrained by the 3-to-1 intensity ratio (and same width) 
of [N{\sc ii}]. These fits badly failed, although the Lorentzians yielded smaller but yet unacceptably large 
residuals. The ratio of the measured peaks of $\lambda6583$ and $\lambda6548$ in the blend is only 1.8 
to 1 instead of the expected 3-to-1 ratio. This indicates that the flux from an additional component is 
stronger beneath $\lambda6548$ than below $\lambda6583$, which can be relatively easily achieved by the 
contribution of a broad H$\alpha$ component {\em in addition} to narrow H$\alpha$. The wide Lorentz 
wings of a single H$\alpha$ component (allowed to be broader than the [N{\sc ii}] lines) cannot lift 
$\lambda 6548$ sufficiently.

All trial fits with an additional broad component assume it to be at the same central wavelength as the 
narrow H$\alpha$ feature. One broad component, either Gaussian or Lorentzian, does not lead to a 
satisfactory fit if three narrow Gaussians are added, due to their typical `lack of wings'. Solely 
equipped with Gaussians, one would need many more than these four components for an acceptable fit.

\begin{figure}[ht]
\resizebox{\hsize}{!}{\includegraphics[angle=-90]{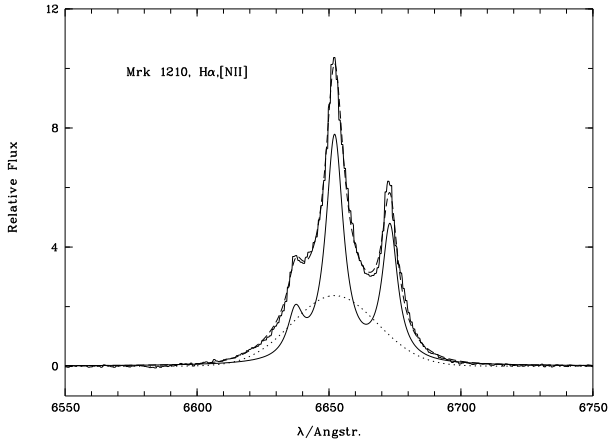}}
\caption{H$\alpha$+[N{\sc ii}]$\lambda\lambda6548, 6583$ from 0\farcs68 east of the region where lines are at maximum. 
The total fit (dashed line) consisting of a sum of three Lorentzians (continuous line) for narrow H$\alpha$ 
and [N{\sc ii}] (the [N{\sc ii}] lines are constrained by the 3:1-ratio and the same width) plus a broad Gaussian (dotted) 
concentric with the narrow Lorentzian that fits the core of H$\alpha$.}
\label{mkha}
\end{figure}

Three narrow and one broad Lorentzian function produce too much flux in the outskirts of the blend 
suggesting that the real underlying broad component does not have extended wings. Indeed, the best four-line 
fits were obtained with three narrow Lorentzians and one broad Gaussian concentric with the narrow Lorentzian 
of H$\alpha$ (as shown in Fig.\,\ref{mkha}). 

Qualitatively, the presence of a broad H$\alpha$ component is not unexpected in view of the polarized (degree 
of polarization $\sim 1\%$) broad H$\alpha$ found in Mrk\,1210 (Tran 1995). However, the faint polarized 
flux should not give such a strong signal in total flux. 
%
%
%

The good single-component fits of [N{\sc ii}] suggest to fit the other strong forbidden-systems by single 
Lorentzian functions as well (with the joint constraints as given in Appendix\,\ref{anal}). The resulting
velocities and line widths from these single-component fits are given in Table\,\ref{mks}. In general, 
these simple fits reproduce the gross structures relatively well and may be considered as a success for 
the Lorentzian functions because {\em single Gaussians fail to represent the wings of all major lines}. 

\begin{figure}[ht]
\resizebox{\hsize}{!}{\includegraphics[angle=-90]{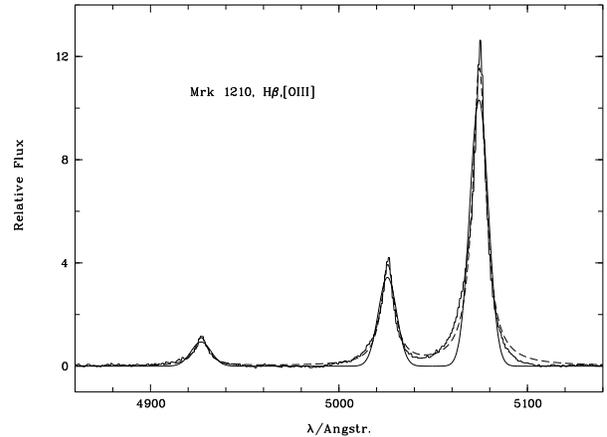}}
\caption{As in Fig.\,\ref{mkha}, but for H$\beta$ and [O{\sc iii}]$\lambda\lambda4959, 5007$ as fit by 
single Lorentzian functions (dashed). The [O{\sc iii}] lines are constrained by the 3:1-ratio and equal width. 
Obviously they have a blueward wing asymmetry. The smooth continuous line shows the analogous, but 
much inferior fit with Gaussians.}
\label{mko3}
\end{figure}

Even H$\beta$ can be represented by a single Lorentzian, somewhat wider than the above fit of narrow 
H$\alpha$ and much narrower than broad H$\alpha$ (Fig.\,\ref{mko3}, Table\,\ref{mks}). This means 
that with the chosen fitting procedure the profiles of H$\alpha$ and H$\beta$ do not agree. The 
broad+narrow component decomposition of H$\alpha$~cannot be transferred to H$\beta$ by allowing for 
different values of extinction for the components.

\begin{figure}[ht]
\resizebox{\hsize}{!}{\includegraphics[angle=-90]{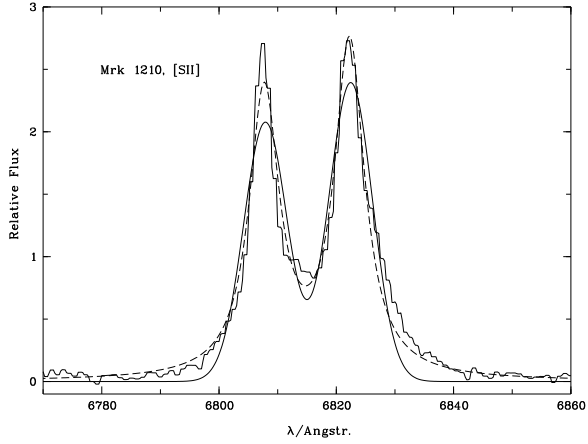}}
\caption{[S{\sc ii}]$\lambda\lambda6716, 6731$ from the same region as in Fig.\,\ref{mkha} as fit by two 
Lorentzians (dashed) with same width and position in velocity space. A fit with two Gaussians and 
otherwise same presumptions is displayed by the smooth continuous line. Athough not perfect, single 
Lorentzians appear to be better suited as basic fit functions than single Gaussians.}
\label{mks2}
\end{figure}

\begin{figure}[ht]
\resizebox{\hsize}{!}{\includegraphics[angle=-90]{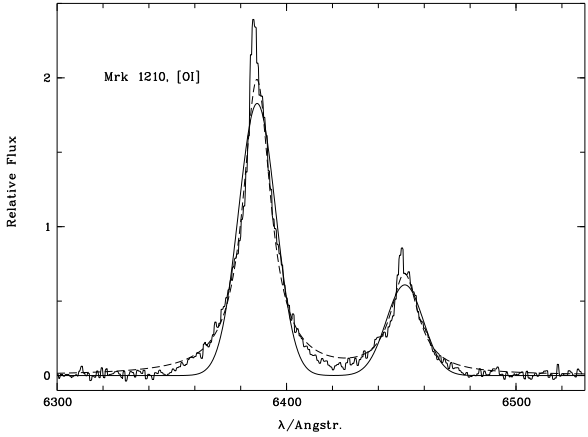}}
\caption{As in Fig.\,\ref{mkha}, but for [O{\sc i}]$\lambda\lambda6300, 6363$ as fit by two Lorentzians 
(dashed) with same width and position in velocity space. Smooth continuous line: The analogous 
trial fit with Gaussians.}
\label{mko1}
\end{figure}

A close look unveils marked imperfections in the forbidden-line fits as well. E.g., the [O{\sc iii}] lines 
are asymmetric to the blue (Fig.\,\ref{mko3}), in the [S{\sc ii}] blend a residual excess remains in the 
red wing (Fig.\,\ref{mks2}). [O{\sc i}] shows structure and is redshifted relative to the other lines 
(Fig.\,\ref{mko1}, Table\,\ref{mks}). In addition, the [O{\sc iii}] and [O{\sc i}] lines ($w \sim 600$\,km/s) 
are significantly wider than the [N{\sc ii}] and [S{\sc ii}] lines ($w \sim 300$\,km/s). Altogether, a more 
sophisticated decomposition is required which will be proposed in the next subsection.
%
%
%
%
%
%
%
%
%
%
%
\begin{table}
\caption[]{Relative intensity ratios of components $\cal{M}$, $\cal{R}$ and $\cal{B}$ as defined by
corresponding velocity-width parameters $(v,w)(\cal{M})$ = (4064, 208), $(v,w)(\cal{R})$ = (4304, 
350) and $(v,w)(\cal{B})$ = (3895, 727) (in km/s). The intensity of the line-core component 
$\cal{M}$ is set to unity. (1) and (2) correspond to alternative fits of the H$\alpha$+[N{\sc ii}] blend 
(Figs.\,\ref{mhal1} and \ref{mhal2}), respectively. }
\label{mt1}
\begin{flushleft}
\begin{tabular}{lcccc}
Line    & $\cal{R}$(1) &  $\cal{B}$(1) & $\cal{R}$(2) & $\cal{B}$(2) \\
\hline\hline
$[{\rm O}${\sc i}]                   & 1.65 &  2.05  &      &       \\
$[{\rm S}$\,{\sc ii}]$\lambda6716$   & 0.18 &  0.07  &      &       \\
$[{\rm S}$\,{\sc ii}]$\lambda6731$   & 0.41 &  0.17  &      &       \\
$[{\rm N}$\,{\sc ii}]                & 0.42 &  1.74  & 0.29 & 0.90  \\
H$\alpha$                            & 0.42 &  1.73  & 0.76 & 1.71  \\
H$\beta$                             & 0.65 &  2.24  &      &       \\
$[{\rm O}$\,{\sc iii}]               & 0.44 &  1.97  &      &       \\
\hline
\end{tabular}
\end{flushleft}
\end{table}
\subsubsection{The decomposition of the line profiles}
Table\,\ref{mks} shows that the velocities of the single-component fits for emission from between 
0\farcs98 east and 1\farcs06 west agree within 10\,km/s so that we combined the lines to represent
the total emission from the $\pm 1\arcsec$ of the brightness maximum. This gives practically no
difference to fits of single extraction-rows, but reduces the noise and is more appropriate in view of
the width of the spatial point-spread function of $\sim2 \arcsec$. 

The red wings seen in the [S{\sc ii}] and [O{\sc i}] profiles and the large widths of [O{\sc i}], 
H$\beta$ and [O{\sc iii}] suggest the presence of different kinematical components. However, the 
H$\beta$ and [O{\sc iii}] profiles and the H$\alpha$+[N{\sc ii}] blend are lacking clear substructure 
for providing anchoring features to constrain different components. At first, we therefore looked 
for guidance by visible structures in other lines and attempted to apply the resulting components 
with the {\em same} widths and positions to the smoother profiles. 

\begin{figure}[ht]
\resizebox{\hsize}{!}{\includegraphics[angle=-90]{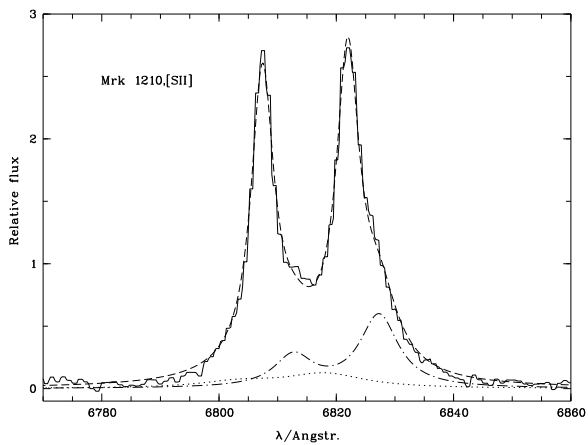}}
\caption{ [S{\sc ii}]$\lambda\lambda6716,6731$ from the central 2\arcsec~of Mrk\,1210 with maximum
line emission as fit by three kinematically separated Lorentzian emission systems: a blue component 
$\cal{B}$ (dotted), which is unusually faint in [S{\sc ii}]; a red component $\cal{R}$ (dash-dotted) 
(explaining the redward asymmetry of [S{\sc ii}]; the main component $\cal{M}$ that fills the cores 
of the lines (not shown)). The dashed line shows the total fit made up of  $\cal{M}$ + $\cal{B}$  
+  $\cal{R}$.}
\label{ms2}
\end{figure}

Fig.\,\ref{ms2} shows a three-component Lorentzian fit to [S{\sc ii}]. $\cal{M}$ (not displayed) is the 
`main' component determined by the peaks of the profiles, which therefore has the closest relation to the
velocity curve in Fig.\,\ref{vmk}. The red-shifted component $\cal{R}$ has been invoked to fit a 
residual shoulder on the red side and the inter-line regime. The relative strength of the two [S{\sc ii}] 
lines of $\cal{M}$ is a free parameter. As in the case of NGC\,1052, we assume that the particular 
strength of [O{\sc i}] is partially due to the second component (here: $\cal{R}$) with $n \gg 10^3$ cm$^{-3}$, 
which fixes its intensity ratio [S{\sc ii}]$\lambda6731$/[S{\sc ii}]$\lambda6716=2.3$. The components are
$\cal{M}$ (`main') and $\cal{R}$ (`red') defined by ($v$=heliocentric velocity, $w$=FWHM), both measured 
in km/s. The fit gives $(v,w)(\cal{M})$ = (4067, 208) and $(v,w)(\cal{R})$ = (4304, 350). The weak 
third component $\cal{B}$ slightly improves the fit of [S{\sc ii}], but its presence is inferred from 
[O{\sc i}] and [O{\sc iii}].

\begin{figure}[ht]
\resizebox{\hsize}{!}{\includegraphics[angle=-90]{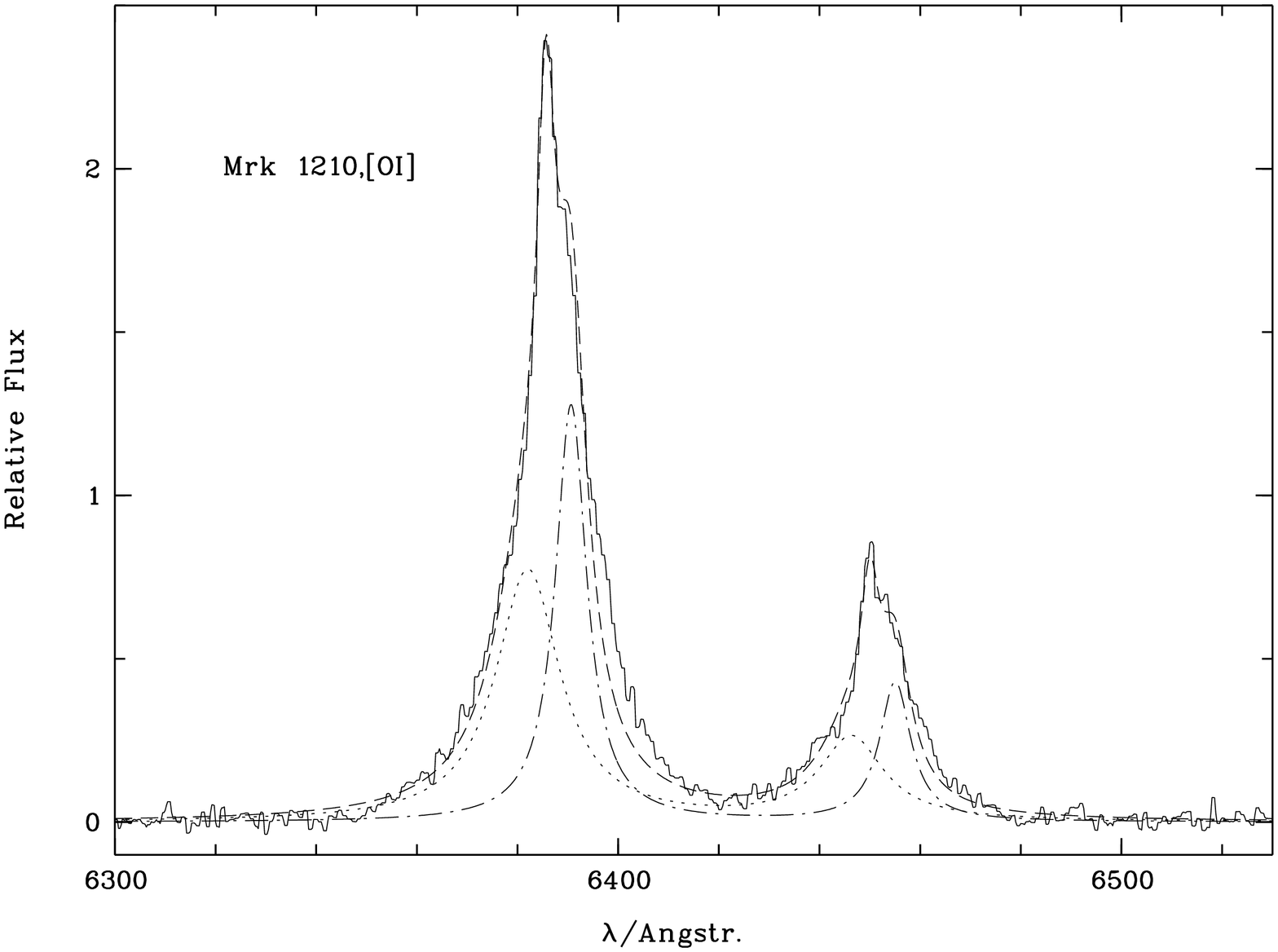}}
\caption{ As in Fig.\,\ref{ms2}, but for [O{\sc i}]$\lambda\lambda6300,6363$ from the maximal 
line-emitting 2\arcsec~of Mrk\,1210. }
\label{mo1}
\end{figure}

\begin{figure}[ht]
\resizebox{\hsize}{!}{\includegraphics[angle=-90]{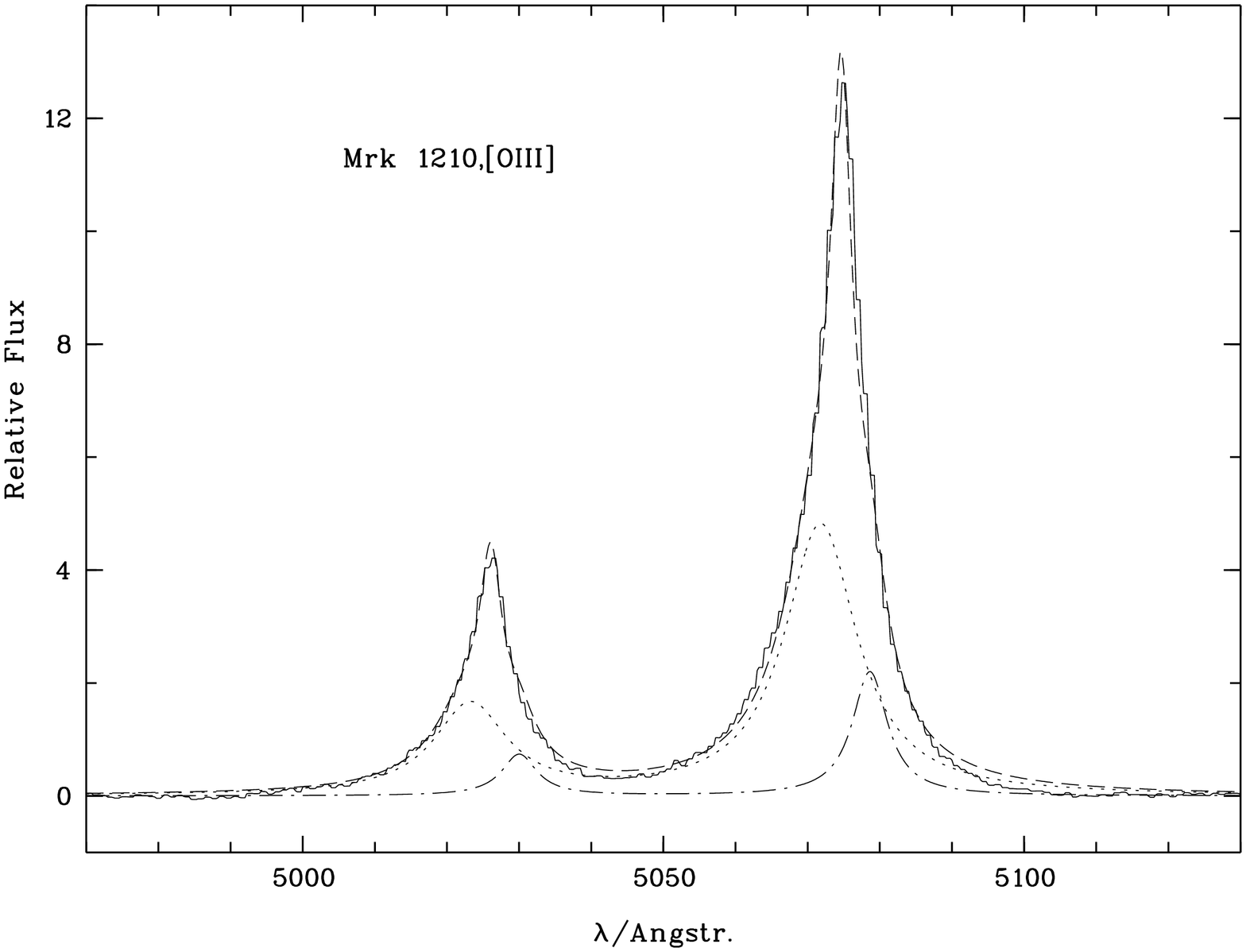}}
\caption{As in Fig.\,\ref{ms2}, but for [O{\sc iii}]$\lambda\lambda4959,5007$ from the maximal 
line-emitting 2\arcsec~of Mrk\,1210.}
\label{mo3}
\end{figure}

\begin{figure}[ht]
\resizebox{\hsize}{!}{\includegraphics[angle=-90]{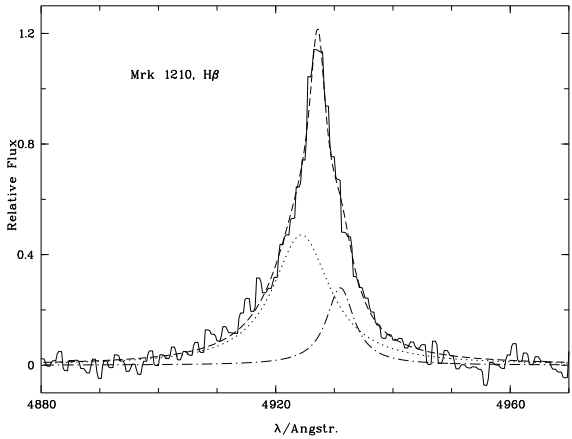}}
\caption{As in Fig.\,\ref{ms2}, but for H$\beta$ from the maximal 
line-emitting 2\arcsec~of Mrk\,1210.     }
\label{mhb}
\end{figure}

Applying $\cal{M}$ and $\cal{R}$ to [O{\sc i}] and [O{\sc iii}] takes care of the shoulder in the upper red 
wing of [O{\sc i}] and yields an excellent fit of the red wing of [O{\sc iii}], but, for the complete profiles, 
an additional `blue' component $\cal{B}$ = (3895, 727) is also required. $\cal{B}$ is an 
excellent supplement to fit both the blue parts of [O{\sc i}] an [O{\sc iii}] (Figs.\,\ref{mo1} and \ref{mo3}). 
Only the amplitudes of the components are left as free parameters to be determined during the fitting 
procedure; widths and positions are the same in all lines. Also H$\beta$, ten times fainter than 
[O{\sc iii}]$\lambda5007$ and therefore noisier, could be easily fit by a sum of $\cal{M}$, $\cal{B}$ and 
$\cal{R}$ (Fig.\,\ref{mhb}).

\begin{figure}[ht]
\resizebox{\hsize}{!}{\includegraphics[angle=-90]{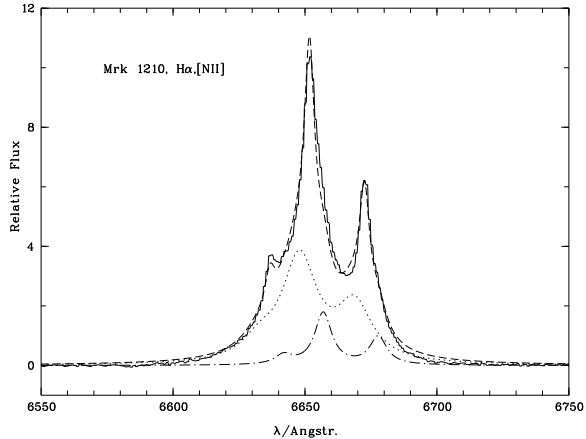}}
\caption{As in Fig.\,\ref{ms2}, but for H$\alpha$+[N{\sc ii}] from the maximal 
line-emitting 2\arcsec~of Mrk\,1210. In this special fit, the [N{\sc ii}]/H$\alpha$ ratio is the same for systems
$\cal{M}$, $\cal{B}$ and $\cal{R}$ (solution (1), see text for more details).}
\label{mhal1}
\end{figure}

\begin{figure}[ht]
\resizebox{\hsize}{!}{\includegraphics[angle=-90]{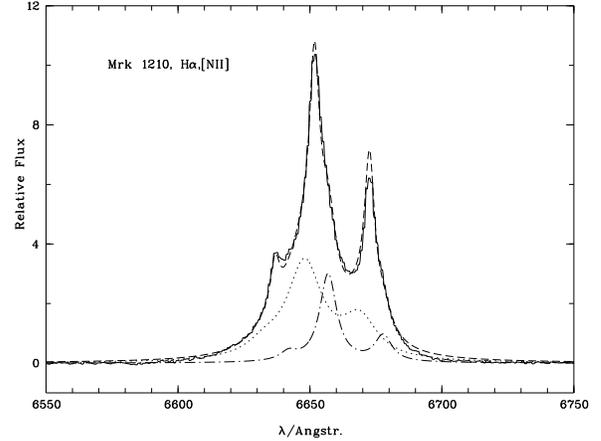}}
\caption{As in  Fig.\,\ref{mhal1}, but here the intensity ratios [N{\sc ii}]/H$\alpha$ are allowed to
differ among systems $\cal{M}$, $\cal{B}$ and $\cal{R}$ (solution (2), see text for more details). }
\label{mhal2}
\end{figure}

To find a corresponding fit of H$\alpha$+[N{\sc ii}] with six free parameters (the amplitudes of H$\alpha$ 
and [N{\sc ii}] for each system) turned out to be too demanding for the software. Setting manual constraints
while guided by the amplitude ratios of the other fits and the response to parameter variations led 
to two acceptable fits, dubbed solution (1) and (2) (Figs.\,\ref{mhal1} and \ref{mhal2}; Tables\,\ref{mt1}
and \ref{mt2}). 

In Fig.\,\ref{mhal1} the main-component amplitudes $h(\cal{M})$ of H$\alpha$ and [N{\sc ii}]$\lambda 6548$ 
are left free, while, in a series of fits, the corresponding $h(\cal{R})$ and $h(\cal{B})$ were 
manually adjusted, which are, however, constrained to have {\em fixed ratios} to $h(\cal{M})$. With 
this procedure each component is confined to the same [N{\sc ii}]/H$\alpha$ ratio. We obtain 
[N{\sc ii}]$\lambda 6583$/H$\alpha= 0.538$ and $h(\cal{M})$:$h(\cal{R})$:$h({\cal B})=1 : 0.25 : 0.50 $ 
(solution (1)).

However, to fix [N{\sc ii}]$\lambda 6583$/H$\alpha$ for all components is a much too simplifying assumption
since it depends on density and input flux in a more complicated manner than, e.g., the Balmer normalized 
intensities of [O{\sc i}] or [O{\sc iii}] (e.g. Komossa \& Schulz 1997). [N{\sc ii}]/H$\alpha$ might differ 
from component to component. Therefore, in another series of fits different values of the relative 
[N{\sc ii}] amplitudes $h(\cal{R})$:$h(\cal{M}$) and $h(\cal{B})$:$h(\cal{M}$) were predefined while 
the H$\alpha$ amplitudes of the components were left as free fit parameters. The best fit (solution (2)) 
is shown in Fig.\,\ref{mhal2}. The resulting intensity ratios are given in Tables \ref{mt1} and \ref{mt2}. 

Due to a lower $\chi^2$ we have a formal reason to prefer solution (2). The second, more convincing,
reason is the more plausible physical assumption underlying fit (2). To achieve uniqueness in fitting 
the H$\alpha$+[N{\sc ii}] blend more constraints than just fitting profiles have to be taken into account.

In solution (2) [N{\sc ii}]$\lambda 6583$ is slightly above the observed level. The discrepancy cannot 
be removed with redward-asymmetric components. In this regard, a broad component would be more 
flexible (see Fig.\,\ref{mkha}). Given the clear evidence for broad H$\alpha$ in {\em polarized} 
flux (Tran 1995), a faint broad component underlies the lines in {\em total} flux as well. The 
fits shown here nevertheless demonstrate that it is not necessary to postulate a broad component.

Table \ref{mt2} gives the major line-intensity ratios for the adopted decomposition, which all lie in 
the AGN regime of the standard diagnostic diagrams of VO.  The [O{\sc iii}]$\lambda5007$/H$\beta$~ratios 
lie in the range $10\pm2$.  $\cal{R}$ is characterized by relatively strong [O{\sc i}]/H$\alpha$ combined 
with relatively weak [N{\sc ii}]/H$\alpha$ and [S{\sc ii}]/H$\alpha$. 
 
We carried out further experiments to check the uniqueness of the fits. It turned out, that the 
visible narrow cores of the lines constrain $\cal{M}$ rather well, and next comes $\cal{R}$ determined 
to fill in the shoulders in [S{\sc ii}] and [O{\sc i}]. The weakest constraint is for $\cal{B}$, which 
takes account `for the rest' and could be widened to $\sim 900$\,km/s (and correspondingly shifted) 
on the expense of the other components. However, experiments in this direction yielded worse total 
fits so that for the final discussion solution (2) given above is preferred.

%
\begin{table*}
\caption[]{Logarithmic intensity ratios as used in the diagnostic diagrams by Veilleux \& 
Osterbrock (1987; VO) of components $\cal{M}$, $\cal{R}$ and $\cal{B}$ of Mrk\,1210. (1) and 
(2) correspond to alternative fits of the H$\alpha$+[N{\sc ii}] blend (Figs.\,\ref{mhal1} and \ref{mhal2}). }
\label{mt2}
\begin{flushleft}
\begin{tabular}{lcccccc}
Line    & $\cal{M}$(1) & $\cal{R}$(1) &  $\cal{B}$(1) & $\cal{M}$(2) & $\cal{R}$(2) & $\cal{B}$(2)  \\
\hline\hline
$\log \, ([{\rm O}$\,{\sc iii}]$\lambda5007$/H$\beta$) &  1.075 &  0.906 &  1.019 &  1.075 &  0.906 &  1.019\\
$\log \, ([{\rm N}$\,{\sc ii}]$\lambda6583$/H$\alpha$) &--0.269 &--0.269 &--0.270 &--0.107 &--0.529 &--0.385\\
$\log \, ([{\rm S}$\,{\sc ii}]$\lambda6716$/H$\alpha$) &--0.515 &--0.879 &--1.883 &--0.490 &--1.109 &--1.854\\
$\log \, ([{\rm S}$\,{\sc ii}]$\lambda6731$/H$\alpha$) &--0.504 &--0.518 &--1.523 &--0.480 &--0.747 &--1.491\\
$\log \, ([{\rm S}$\,{\sc ii}]$\lambda\lambda 6716, 6731$/H$\alpha$) 
                                                       &-0.211  &--0.361 &--1.366 &--0.184 &--0.591 &--1.334\\
$\log \, ([{\rm O}$\,{\sc i}]$\lambda6300$/H$\alpha$)  &--0.647 &--0.055 &--0.574 &--0.343 &--0.285 &--0.543\\
\hline
\end{tabular}
\end{flushleft}
\end{table*}

%
%
%
%
%
%
%
%
%
%
%
%
%
%
%
%
%
%
%
%
\section{Discussion}
\label{disc}
\subsection{The narrow-line profiles of NGC\,1052 and Mrk\,1210}
\subsubsection{Kinematic bulk components}
\label{kin}
The slightly novel {\em ansatz} of the present work lies in the attempt to disentangle two 
to three emission-line systems from the observed line profiles. {\em Ab initio} such a task 
may be subject to pitfalls like non-uniqueness and particular care has to be devoted to the 
choice of a basic function that represents a single profile.

The above fitting experiments carried out with the spectra of NGC\,1052 and Mrk\,1210 showed 
that single-component fits as well as multi-component fits led to lower residuals and a lower 
number of fitting components when using Lorentzians as basic functions rather than the commonly 
employed Gaussians. The additional surprising effect of this choice was that we could essentially 
dispense with broad H$\alpha$ (and H$\beta$) components. There is no doubt that both galaxies 
display broad H$\alpha$ components in polarized flux but these are faint and are unlikely to be 
visible in spectra taken without a polarizing device.

If the Lorentzian decomposition is correct, the broad H$\alpha$ component extracted by Ho et al.
(1997) in the total-flux spectrum of NGC\,1052 has to be considered as spurious. Our alternative 
decomposition at least suggests to be cautious with claiming a broad H$\alpha$ component hidden 
in the H$\alpha$+[N{\sc ii}] blend if it is only based on a mere decomposition in total flux. In this 
view it seems to be a coincidence that spectropolarimetry finally confirmed NGC\,1052 as a LINER-1.

Before giving possible astrophysical settings for Lorentzian profiles, an empirical justification 
is in order. The significance of the decomposition into different Lorentzian components rests on the
internal consistency achieved for all strong line profiles from H$\beta$ to [S{\sc ii}]$\lambda6731$. 
In particular, the same decomposition was employed for the collisionally excited transitions
of [O{\sc i}], [O{\sc iii}], [N{\sc ii}] and [S{\sc ii}] as for the allowed recombination lines 
H$\alpha$ and H$\beta$. Because of electron densities believed to exceed $10^9$\,cm$^{-3}$, a classical 
AGN BLR would only exhibit {\em broad allowed} lines (with line widths (FWHM) in the range 
$10^{3-4}$\,km/s) so that we consider the components extracted here (with FWHM $\sim 10^{2-3}$\,km/s) 
as belonging to parts of a NLR distinguished by different bulk motions. 

Density tracers usually suggest densities in the range $10^{2-4}$\, cm$^{-3}$ for the main components 
of NLRs. For NGC\,1052 and Mrk\,1210 blue components $\cal{B}$ with indications for rather high 
densities ($\sim10^6$\,cm$^{-3}$ or more) were isolated.  In particular, $\cal{B}$ of Mrk\,1210 
hardly shows up in [S{\sc ii}]$\lambda\lambda6716,6731$, which suggests $n_e \gg 10^3$\,cm$^{-3}$. However, 
since the critical densities of the well observed doublet [N{\sc ii}]$\lambda\lambda6548, 6583$ are 
$\sim 10^5$\,cm$^{-3}$, the latter value should not be exceeded by several orders of magnitude
Furthermore, the trends shown in Komossa \& Schulz (1997) indicate, that to boost [N{\sc ii}]/H$\alpha$ 
relative to [S{\sc ii}]/H$\alpha$ at high densities, a large ionizing input flux is necessary. $\cal{B}$ 
might therefore be an `ILR' (intermediate-line region) with a density and a distance intermediate 
between the BLR and NLR. This designation is also supported by its large line width of 727\,km/s, 
somewhat intermediate between a `typical' NLR and a typical BLR.

In the early days of NLR analysis width-variations of narrow lines were frequently attributed to
stratifications in velocity space observationally manifested via different critical densities or
ionization potential (e.g. Pelat et al. 1981, Filippenko \& Halpern 1984, Whittle 1985b, Schulz 1987).
The decompositions proposed here explain different line widths by summing line components, each
of {\em constant} width in velocity space, but different weighting for a component in each line. 
This assumption implies that each emission-line cloud emits all lines considered here, which
will be the case when they are mainly ionization bounded.

While our data suggest that the $\cal{M}$ components arise in a rotating disk with `normal' 
densities (their velocities fit the rotation curves), the $\cal{R}$ and $\cal{B}$ components appear 
to arise in bulk outflow systems. 
%
%
\begin{figure}[ht]
\resizebox{\hsize}{!}{\includegraphics[angle=-90]{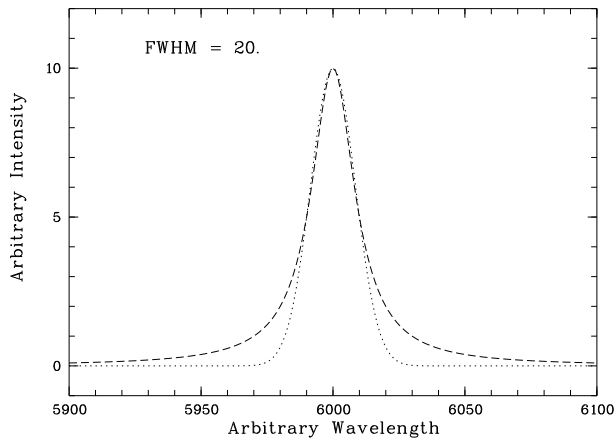}}
\caption{A Lorentzian (dashed) and a Gaussian (dotted) with same full width at half maximum.}
\label{lg}
\end{figure}
\subsubsection{Why Lorentzians?}
\label{sca}
Gaussians are relatively well suited to describe instrumental functions or thermal Doppler 
broadening or some kind of turbulence induced by random motions with Maxwellian-type velocity 
distributions. Lorentzians differ from Gaussians with same FWHM by a marginally narrower core 
and a slower fall-off to large velocities leading to more extended line wings (Fig.\,\ref{lg}).
The non-Gaussian nature of {\em observed} [O{\sc iii}] line profiles was early described by Whittle 
(1985a), who had noted that (i) ``almost all the sample have [O{\sc iii}] profiles with a stronger 
base relative to the core than Gaussian'' (his Sect.\,5.3) and (ii) ``the extreme wing widths (FWZI) 
usually lie in the range 1000--3000\,km s$^{-1}$'' (his Sect.\,5.2). 

For our line profiles, the observed line wings would have typically reached the noise levels at 
$\sim 6\%$ of the maximum. Eq.\,\ref{gaulo} implies that a Gaussian falls to 6.3\% of the maximum 
at two HWHMs (half width at half maximum) from the center while a Lorentzian has decreased to 5.9\% 
at four HWHMs. For components $\cal{M}$, $\cal{R}$ and $\cal{B}$ of Mrk\,1210 described in 
Table\,\ref{mt1} we get 4 HWHMs = $2\,w = 416, 700, 1454$\,km/s, respectively. How can such large 
velocities of $\sim 10^{2-3}$\,km/s in the line wings be attained?

Turbulence of this order {\em within} the emission-line clouds would be strongly supersonic in gas 
with an observed temperature of $T \sim 10^4$ K, in which the characteristic thermal velocity and 
the velocity of sound are of the order of $\sim 10$\,km/s. Hydrodynamic shocks would result in 
dissipation of the gained energy, most likely by radiation, perhaps contributing to, but not 
dominating the ionization equilibrium that is essentially governed by radiative input. To uphold 
the velocities, a strong steady mechanical energy input would be required. 

Outflow velocities of the order of $10^3$\,km/s have frequently been seen in AGN absorption lines. 
Elvis (2000) reviewed the observations and proposed an AGN-type-I unification scheme. Cloudlets
comoving in a bipolar wind system might lead to such line wings whose width would strongly depend 
on the viewing angle of the system.  

A less explored, albeit not far-fetched, mechanism to produce the line wings might be magneto-hydrodynamic 
waves that could mimick supersonic turbulence (Shu 1992).  

A different type of alternative would be scattering of photons at fast particles. Dust particles are 
far too slow, but electrons in the cited photoionized clouds have velocities of $10^{3-4}$\,km/s. 
Scattering of a line at these electrons could cause broad line wings via redistribution in frequency 
space. In the early times of AGN spectroscopy, this type of scattering was proposed to explain the 
widths of lines from the BLR (Weymann 1970; Mathis 1970). In a hotter intercloud medium with $T 
\sim 10^{6-8}$ K, which may confine the emission-line clouds by its thermal pressure, electrons can 
attain velocities of $10^{4-5}$\,km/s. Shields \& McKee (1981) suggested electron scattering in such 
a medium as an explanation for very broad H$\alpha$ wings observed in the quasar B\,340. Their case 
could be transferred to the Lorentzian wings, if one sees only the `tip of the iceberg' of hot-ISM 
scattering because otherwise the wings would have to be much wider.

The fraction of scattered flux is roughly given by the electron-scattering optical depth $\tau_e$ 
and the covering fraction of the scattering medium. Effective column-densities of fully ionized gas 
of $2\,10^{22}$ cm$^{-2}$ would be required to obtain electron-scattering optical depths on the percent 
level. Attributing such a column to a kpc-extended sphere of a hot intercloud medium would lead to a 
total bremsstrahlung luminosity (Eq.\,5.15b in Rybicki \& Lightman 1979) of $2.6\,10^{42} 
(n_e/10\,{\rm cm}^{-3})^2 \sqrt{(T/10^6 {\rm K})} (R/500 {\rm pc})^3$ erg/s. This is a lower limit 
because the estimate neglects line emission. Thermal soft-X-ray emission is estimated to be $\sim 10^{40}$ 
erg/s for NGC\,1052 (Weaver et al. 1999), and is measured to be $\sim 10^{41}$ erg/s in Mrk\,1210 
(Awaki et al. 2000). We conclude that we do not see a sufficient amount of hot coronal gas to produce 
the required wings. However, a viable alternative would be scattering at the cooler ($T\sim 10^4$ K) 
emission-line clouds with electron densities $\sim10^{2-3}$. With a scale $R\approx 10^2$ pc and a 
geometrical scattering factor $f\approx0.3$ one easily obtains a ratio of scattered luminosity to 
input luminosity $L_{\rm s}/L_{\rm in} = f \tau_e$ of a few percent, which suffices to explain th
extended wings of the narrow lines. 
 
\subsubsection{Relationship to spectropolarimetry}
\label{pol}
Spectropolarimetry of NGC\,1052 (Barth et al. 1999) and of Mrk\,1210 (e.g. Tran 1995) provides direct
evidence for scattering of the continuum and broad lines. 

In Mrk\,1210 the degree of polarization $P$ in the forbidden lines decreases significantly (to less
than 1\%; to $\sim 0$\% in the cores of [O{\sc iii}]) compared to that of the continuum indicating that 
the main body of the {\em narrow-line profiles} is {\em not scattered} (this is the situation in 
many `hidden-Seyfert-1' galaxies, which are thereby backing the unified model). In NGC\,1052 there 
is hardly a decrease in the forbidden lines, suggesting that a major part of the forbidden-line 
profiles is scattered as well, rendering this object as an unusual `hidden-LINER-1' object. How
can these observations be reconciled with the line-profile decomposition proposed here? We discuss
the galaxies in turn.

The average angle of polarization in Mrk\,1210 is $\sim 30\degr$ (Tran 1995). This is 80\degr~off 
from the kinematical major axis of the central kpc-disk (cf. Table\,\ref{t2}), whose geometry is 
not favorable to give rise to polarized radiation because of its near-face-on view. A suggestive 
explanation is found by the observed shift of the line-continuum maximum by $\sim 1\arcsec$
to the west of the kinematical center (see Fig.\,\ref{vmk}, where the zero point of the abscissa 
is at the brightness maximum), which is consistent with a view that the true AGN is obscured and 
is situated at the kinematical center. Via some escape route AGN radiation reaches the brightness center,
thereby giving rise to the polarized scattered broad-line wings and the polarized part of the continuum.
In the AGN standard view, the high-excitation Seyfert-2 characteristics of the brightness center
would be traced back to photoionizing AGN radiation reaching this region. Because the AGN continuum 
is only seen in scattered light and the stellar continua from the starburst plus the older stellar 
populations certainly dominate, it is not surprising that the precise amount of visible `featureless 
continuum' is under debate (Schmitt et al. 1999). 

In NGC\,1052, Barth et al. (1999) found a polarimetric position angle of H$\alpha$ (+[N{\sc ii}]) of 
178\degr, which is neither aligned with -- nor perpendicular to -- the major axis of the gaseous disk
(p.a.\,$\sim45\degr$~following Davies \& Illingworth (1986)). However, this apparently accreted 
narrow-line gas does not necessarily reflect the orientation of the `central machine'. If the 
kpc-radio structure detected by Wrobel (1984) can be identified with a jet perpendicular to the 
central accretion disk, the polarimetric p.a. would perfectly fit with the idea that the polarized 
component is scattered light originating in a nucleus obscured from direct view.

However, unlike in Mrk\,1210 (Tran 1995; second panel in his Fig.\,15), in NGC\,1052 the degree of 
polarization $P$ (Barth et al. (1999); their Fig.\,1, middle and lower panel, and Fig.\,2, lower panel) 
only hardly decreases in the [N{\sc ii}]+H$\alpha$ blend and shows upturns in [O{\sc i}] and [S{\sc ii}]. 
Considerable noise and possible Galactic foreground polarization affecting the spectrum lets Barth et al. 
focus on [O{\sc i}]$\lambda6300$, whose value of $P$ lies significantly above that of the continuum. In 
particular, they note that [O{\sc i}] is narrower in polarized light and that the [O{\sc i}]/[S{\sc ii}] 
flux ratio is larger in Stokes flux than in total flux. 

These latter findings resemble those we found for component $\cal{B}$ in NGC\,1052 (cf. Table\,\ref{ab52}, 
last two rows). Relative to $\cal{A}$, in $\cal{B}$ usually [O{\sc i}] is stronger and [S{\sc ii}] is weaker.
The dense narrow-line gas of $\cal{B}$ could be (partially) hidden and its radiation be scattered
like the broad polarized wings of H$\alpha$.  

Alternatively, Barth et al. (1999) suggested transmission through a region of aligned dust grains as a
possible mechanism for the [O{\sc i}] polarization. Indeed, Davies \& Illingworth (1986) noted a dusty zone
on the eastern side of the nucleus where component $\cal{B}$ is predominantly detected. 
\subsection{Relationship to H$_2$O megamaser activity}
\subsubsection{General aspects}
The $\lambda$$\sim$1.3\,cm transition of H$_2$O can be driven to mase by collisions, provided 
that the densities and temperatures are high enough ($\ga$10$^{7}$\,cm$^{-3}$ and $\ga$400\,K; see 
Sect.\,1). In galactic sources of water maser emission, namely in star forming regions, the necessary 
conditions can be produced by shock waves which provide not only proper physical conditions (e.g. Elitzur
1992, 1995) but also greatly enhanced H$_2$O abundances (e.g. Melnick et al. 2000). The unique 
association of H$_2$O megamasers with AGN and the fact that not all of the megamaser emission can 
arise from unsaturated amplification of an intense nuclear radio continuum (e.g. Greenhill et al. 
1995a; AGN brightness temperatures can be much higher than $T_{\rm b}$$\la$10$^{4}$\,K, encountered in 
ultracompact galactic HII regions) indicates, however, significant differences between galactic maser 
and extragalactic megamaser sources. Possible scenarios that may produce suitable physical environments
for luminous megamaser emission from the so-called `disk-masers' (see Sect.\,1) are (1) irradiation 
of dense warped disks by X-rays from the AGN (Neufeld et al. 1994), (2) shocks in spiral density waves 
of a self-gravitating disk (Maoz \& McKee 1998) and (3) viscous dissipation within the accretion disk 
(Desch et al. 1998). For the so-called `jet-masers', interaction of dense neutral gas with a radio jet 
is a viable scenario that may also yield temperatures of several 100\,K and large H$_2$O abundances 
(Peck et al. 2001). 

\subsubsection{Outflow as H$_2$O maser trigger: winds of an AGN or a starburst?} 
In view of the overall angular scales, i.e. a few milliarcseconds for interferometric maps 
of 1.3\,cm megamaser emission and a few arcseconds for our study, a connection between optical 
and radio data is difficult to establish. In elliptical galaxies the large scale dust lanes and 
nuclear jet orientation is correlated (e.g. Kotanyi \& Ekers 1979; M{\"o}llenhoff et al. 1992). 
In Seyfert galaxies, however, large scale disks and nuclear disks tend to be inclined with respect 
to each other (e.g. Ulvestad \& Wilson 1984). In some galaxies, the nuclear jet may be oriented not 
too far from the plane of the parent galaxy and may thus interact with molecular clouds, forming 
jet-masers off the very nuclear region (see e.g. Gallimore et al. 2001 for the case of NGC\,1068). 
The near-to-face-on large scale view onto Mrk\,1210 (Malkan et al. 1998) is thus not arguing 
against high shielding column densities along the line-of-sight toward the nucleus. 

Radio jets may be a strong agent to drive maser action when hitting high-density clumps or
entraining molecular gas (see Elitzur 1995 for physical processes involved). However, optically
detectable ionized outflowing material, e.g. a magnetized wind from an accretion disk (cf. Krolik 
1999) that would be less violent close to the equatorial plane than towards higher latitudes, 
may also be relevant. In systems with a nuclear thin disk and a thicker outer torus or in systems 
with misaligned angular momenta of nuclear and outer disk, a nuclear wind impinging on 
molecular clouds farther out appears to be an attractive alternative to the previously outlined 
jet-maser excitation scenario. As in the case of an interaction with a jet, shocks at speeds not 
destroying the dust grains and a warm (several 100\,K) H$_2$O enriched postshock medium are expected. 
The spatial extent of the outflowing gas that may have a larger `cross section' than the faster radio
jet makes such an interaction particularly likely. 

If one relies on outflow as one of the primary maser agents, it must be intriguing that 
H$_2$O megamasers have, until recently, only been detected in galaxies with AGN-typical features 
because large-scale winds and outflows are properties not only associated with AGN but also with 
starbursts. The key to this conundrum may be the fact that the sample of known H$_2$O megamaser sources 
is still rather small. Recently H$_2$O megamasers have been observed, for the first time, in starburst 
galaxies (Hagiwara et al. 2002; Peck et al. 2003) and for the `traditional' absence of H$_2$O megamasers 
in type-1 AGN, proposed by Braatz et al. (1997), a first counterexample may have been found as well 
(see Nagar et al. 2002).

Below, some characteristic features of our four individual galaxies are briefly summarized. 

\subsubsection{Individual sources} 
In {\em NGC\,1052}, a notable feature is the presence of a strong high-density outflow  
(Sect.\,4.3.2). For this galaxy the (VLBA+VLA) observed maser sources are found to coincide with 
the innermost part of the southwestern radio jet rather than being related to the nearby
putative nuclear disk (Claussen et al. 1998). The masers are redshifted by 80 to 200\,km/s 
relative to $v_{\rm sys}$ as tabulated by de Vaucouleurs et al. (1991). The distance to the 43\,GHz 
radio core, the supposed location of the central engine, is less than a 
milliarcsecond (0.07 pc). 
 
Due to the spatial coincidence Claussen et al. (1998) suggested to relate the maser activity to the 
VLBI jet or to a fortuitously positioned foreground cloud that amplifies the radio continuum
of the inner jet. The strong high-density outflow disentangled from the rotating disk (Sect.\.4.3.2)
raises another alternative source to excite the maser activity. In this case the small solid angle 
of the inner jet, the presumably larger solid angle of the outflowing component and the location of the 
masers in front of the brightest parts of the jet suggest unsaturated H$_2$O emission amplifying the 
strong radio background. 

Towards {\em Mrk\,1210}, the brightness maximum BM of the emission-line region is rather compact 
(Falcke et al. 1998) and BM is slightly misplaced from the kinematical center. Strong radial flows in BM 
may be related to the flows that give rise to megamaser activity. No interferometric radio map of the 
H$_2$O emission has yet been published.

In {\em IC\,2560} and {\em NGC\,1386} we found little obvious evidence for outflow, presumably because of the 
near-edge-on view on these galaxies making a flow perpendicular to a disk hard to detect. The NGC\,1386 
line profiles show pronounced structure, and Rossa et al. (2000) isolated even more subtle substructures 
suggestive of non-circular motions like locally expanding gas systems. Such more localized flows could 
be related to megamaser activity as well. However, we caution that integrating over l.o.s., which intersect 
a dusty and patchy rotating disk, can well lead to bumpy line profiles. Interpreting such bumps by local 
Gaussians may be misleading. A meaningful discussion requires observations with significantly higher 
resolution and a detailed model of the dusty spiral disk. While the megamaser in IC\,2560 may originate 
from a circumnuclear disk not related to outflowing gas (Ishihara et al. 2001), no high resolution H$_2$O 
map has yet been published from NGC\,1386.

In passing, it should be noted that the mere presence of outflow in H$_2$O megamaser galaxies is not new. 
In much more detail outflow bulk systems were studied in the nearby galaxies M\,51 (Cecil 1988) and NGC\,4258 
(Cecil et al. 2000), but a direct relationship to their megamaser activity (e.g. Herrnstein et al. 1999;
Hagiwara et al. 2001) has so far been elusive. M\,51 and NGC\,4258 are likely not suitable candidates 
to prove such a relationship, but optical data with subarcsecond resolution are needed for a convincing 
demonstration in other sources.

\section{Final conclusions} 
We have analyzed exploratory optical spectra of four H$_2$O megamaser galaxies in order to disentangle 
kinematically discernable bulk components of their emission-line regions. For each object we propose 
a model of the global kinematics. The gas dynamics in IC\,2560 and NGC\,1386 is mainly determined by 
gravitationally dominated motions in a disk. However, there is a bias due to the close-to-edge-on 
view onto their disks making minor-axis outflow hard to detect. For NGC\,1052 and Mrk\,1210 emission 
from gravity-dominated dynamics of orbiting clouds could be separated from that of systems being in 
outflow. Line-broadening within the individual bulk components appears to be characterized by scattering 
wings.

IC\,2560 is a highly inclined Sb spiral with a nuclear high-ionization Seyfert-2 spectrum displaying 
relatively narrow (FWHM $\sim 200$\,km/s on an arcsecond scale) emission lines from a central rotating 
disk. If its major-axis orientation and inclination agree with that of the outer galaxy, the core mass 
(radius 100 pc) will be $9\,10^6$\,M$_{\odot}$, consistently larger than the $2.8\,10^6$\,M$_{\odot}$,  
given by the orbital velocities of the maser sources presumably embedded in a compact disk of 
outer radius 0.26\,pc as measured by Ishihara et al.\,(2001). A slight blueward asymmetry of the line 
profiles suggests the presence of outflow (Sect.\,4.4.1, Fig.\,3).

The nuclear mass of $\sim 10^7$\,M$_{\odot}$ in IC\,2560 appears to be rather small. It would be larger
by a factor of 1.5 if a distance of 39 Mpc (see comment in Table\,2) were adopted.

NGC\,1386 is a highly inclined Sa spiral with a nuclear high-ionization Seyfert-2 spectrum displaying 
rather structured asymmetric emission-line profiles that we interpret as arising from various parts of 
a central edge-on orientated disk. We dismiss the term `radiation cone' for this object because the
elongated H$\alpha$ morphology may simply reflect the edge-on view on the matter-bounded disk. The 
morphology together with analogy reasoning suggests that the disk is a dusty gaseous mini-spiral, which
appears to be warped. Its rotation curve inside a diameter of 1.6 kpc yields a dynamically acting mass 
of $\sim 5\,10^9$\,M$_{\odot}$, while the visible line-emitting ionized gas has a mass of $\sim 
10^{5-6}$\,M$_{\odot}$, which is a very small fraction of the total mass of nuclear gas that is likely
to be of the order of ten percent of the dynamical mass.

There is a lack of line emission from the kinematical center (the center of the disk), which is probably 
due to obscuration by an equatorial dust disk that is evidenced by a dust lane crossing the center. A 
slight east-west gradient in velocity arising in protrusions seen on an HST image (Ferruit et al. 2000) 
could be due to minor-axis outflow, i.e. a flow essentially perpendicular to the disk (Sect.\,4.2.2).

AGN narrow-line emitting disks like in IC\,2560 and NGC\,1386 explain general findings that the widths of
narrow-lines are broadly correlated with the nuclear stellar velocity dispersion (Nelson \& Whittle 1996).
Such NLRs are likely to be part of the nuclear gas-dust spirals commonly seen on HST images of AGN on a 
scale of $\sim10^{3}$ pc (Malkan et al. 1998; Martini \& Pogge 1999; Martini 2001). However, so far 
unobserved {\em very narrow} lines should be emitted from galaxies with central disks seen face-on, if 
NLRs were {\em only} dominated by disk dynamics. The other two (non-edge-on) galaxies provide examples
for additional dynamical components.

NGC\,1052 is an elliptical galaxy with a nuclear low-ionization emission line region (LINER). All major 
line profiles from the brightness center out to 2\arcsec~northeast can be decomposed into two components 
$\cal A$ and $\cal B$, each with the same width in velocity space for all lines at a given spatial position.  
$\cal A$ is attributed to a rotating disk, while $\cal B$ is blueshifted by $\sim 400$\,km/s suggesting 
outflow of high-density gas relative to the disk. Line ratios of both $\cal A$ and $\cal B$ are LINER-like. 
We propose to relate $\cal B$ to the peculiar forbidden-line polarized component detected by Barth et al. 
(1999) (Sect.\,\ref{pol}). A faint broad H$\alpha$ component, which is seen in polarized flux, is not 
required in total flux. The presence of strong high-density outflow suggests to explore the possibility 
whether such a flow could be an alternate trigger of the H$_2$O megamaser activity in NGC\,1052.

Mrk\,1210 is an amorphously looking Seyfert-2 galaxy that contains a nuclear kiloparsec-spiral seen
face-on. A bright emission-line spot appears to be slightly shifted from the partially obscured kinematical 
center. In total (unpolarized) flux the emission from the bright spot can be decomposed into three 
narrow-line components $\cal M$ (the `normal' main component), $\cal R$ (a redshifted component), and
$\cal B$ (a blueshifted component) without any BLR component. $\cal M$ is tentatively attributed to the 
rotating spiral from which only a small velocity variation could be measured because of its near-face-on 
orientation. $\cal R$ and $\cal B$ are outflow components that may both be related to the cospatial 
starburst and to ionization by the active nucleus. The megamaser activity might be related to the outflow.

The line ratios from the bulk components of Mrk\,1210 fall into the AGN regime of the VO diagrams. 
The shifts of component $\cal{R}$ (by 240\,km/s to the red of $\cal{M}$) and of component $\cal{B}$ 
(by 169\,km/s to the blue relative to $\cal{M}$) are moderate, but of significant influence because 
major portions of the emission-line region are involved. Regarding line widths and densities $\cal{B}$ 
shows characteristics of an intermediate-line-region between BLR and NLR (Sect.\,\ref{kin}).

Finally, it should be noted that the successful line-profile decompositions rest on Lorentzians rather 
than Gaussians as basic functions. Hence, extended wings of {\em intrinsic narrow-line profiles of the 
bulk components} may not be uncommon. Crude estimates suggest that electron scattering at the ionized 
gas itself might lead to a viable explanation, but detailed modelling of such processes would be useful. 

There exists a possible connection between optical and radio lines. It is suggested that nuclear 
outflows impinging onto dense molecular clouds may provide a suitable trigger for megamaser emission.
If inner torus and outer disk are misaligned, such an interaction appears to be particularly likely. 
A convincing verification of such a scenario requires, however, optical spectroscopy with subarcsecond 
resolution.
\appendix
\section{Extraction of emission lines}
A line intensity is obtained by integrating over a line profile in a flux calibrated spectrum. Both 
this integration and any profile decomposition are usually carried out after having subtracted the 
template spectrum and, if this has not led to a `zero base line', a normalizing continuum has been 
subtracted as well. The latter is defined by a spline fit through marks set with the mouse cursor. 
The subjectivity of this eye fit hardly affects our basic results which rely on strong lines. The 
second step is necessary in case of an inappropriate template and/or the presence of a nonstellar 
continuum.

For each object, we define the spatial `center' of an emission line region by the intensity maximum 
of the emission lines. In all cases this center coincides with the maximum of the continuum between 
the lines within a small fraction of an arcsecond. This kind of center may not coincide with the 
kinematical center (usually the 'nucleus' of a galaxy) derived from an assumed symmetry of the velocity 
curve. 

In the main text, discussion of emission-line intensity ratios is largely confined to the diagnostic
classifiers proposed by VO which do not require large and notoriously uncertain reddening corrections. 
Occasionally intensity ratios of the [S{\sc ii}] pair $\lambda6716$ and $\lambda6731$ will be given, which 
may be converted into some kind of weighted mean of the electron densities in the emission-line gas 
via Fig.\,5.3 in Osterbrock (1989).
\section{Common line-profile analysis procedures}
\label{anal}
In AGN even the `narrow' lines show considerable profile structure if measured at sufficient resolution
(Vrtilek \& Carleton 1985; Whittle 1985a). When single-component Gaussian fits fail, commonly 
multi-component Gaussian fits are employed. However, our multi-component fitting experiments of the 
spectra from NGC\,1052 and Mrk\,1210 showed that in these cases Lorentzian functions $C(\Delta \lambda)$ 
(also called `Lorentzians'; $C$ stands for `Cauchy function') appear to be better suited if one intends, 
applying Occam's razor, to minimize the number of required components and parameters in a multi-function fit. 
This may not be generally true. An instructive case of an asymmetric line profile being Lorentzian on 
the blue side and Gaussian on the red side is given by IC\,2560 (cf. Sect.\,4.1).

In a crude picture, the basic kinematical configurations expected are rotating disks, cones of outflowing 
gas or expanding shells. As shown in Schulz et al. (1995), line-of-sight integrations through such 
configurations would neither yield Gaussians nor Lorentzians. However, Schulz et al. (1995) also showed 
that the spatial and spectral smearing by an instrumental function that is wide in comparison to kinematical 
gradients and the presence of strong random motions superposed on the systematic bulk motions will lead 
to profiles characterized by a more symmetric core-wing structure.

This case applies here because our spectral resolution of $\sim 100$\,km/s and spatial resolution of 
$\sim (1-2) \arcsec$~ leads to integration over relatively large cells of the phase space of a NLR.
Hence, for finding bulk motion components experimental fits are carried out by making use of Gaussians 
and Lorentzians as basic functions. 

Lorentzian and Gaussian functions are fixed in the MIDAS data reduction software by the three convenient 
parameters central line position $\lambda_0$ (so that $\Delta \lambda = \lambda - \lambda_0$), $w = 
{\rm FWHM}$ = full width at half-maximum and $h$, the maximal value of the profile function: 
\begin{eqnarray}
\label{gaulo}
C(\Delta \lambda) & = & \frac {h} {1 + 4\, \Delta \lambda^2 / w^2} \\
G(\Delta \lambda) & = & h \exp \left( - \ln (2) \, (4 \Delta \lambda^2 / w^2) \right)
\end{eqnarray}
The ratio of the intensities of lines arising from the same upper level is given by the ratio of the 
corresponding transition probabilities (taken from Tables 3.8 and 3.10 in Osterbrock (1989)). Therefore 
the fits of [O{\sc iii}]$\lambda\lambda4959, 5007$, [O{\sc i}]$\lambda\lambda6363, 6300$ and 
[N{\sc ii}]$\lambda\lambda6548, 6583$ are constrained by fixing intensity ratios in the proportion 1:3 
and by equalizing the FWHM $w$. 

Aside from these constraints, the basic strategy of fitting the line profiles is to start with a minimal 
set of assumptions. At first it is attempted to fit each narrow line by a single Lorentzian function or 
a Gaussian. If this fails we look for features (shoulder, second peak, asymmetric wing) giving clues for 
the initial condition to start a fit with a second component. In case of success, this component (with 
same velocity width $w$, where ($w$ in km/s) = (c in km/s) $\times$ (FWHM in \AA) / ($\lambda_{\rm line}$ in 
\AA)) will be applied to the other strong lines. If the positions among different lines agree sufficiently 
well in velocity space, the fit is considered as a success. In case of Mrk\,1210 a three-component fit 
turns out to be necessary for which the positions of the components are also fixed only leaving the 
component intensity as a free parameter in different lines.

In physical terms, a bulk-motion system would emit lines of approximately the same width if it is, e.g., 
composed of local emitters (with sharp lines compared to the spectral resolution), which are photoionized 
to such a depth that they emit all lines used in the analysis.

Component line widths (FWHM) will be given as measured, i.e. without correcting for our finite spectral
resolution of 100\,km/s. Applying the convolution theorem for two Gaussians, a measured FWHM of 200\,km/s 
would be 14\% too large, with a rapidly decreasing relative excess for larger FWHM (6\% at 300\,km/s;
2\% at 500\,km/s). 

\section{The derivation of rotation curves}
To fix notations and assumptions we briefly describe how rotational velocities are derived for inclined 
plane disks. The disk may recede with a heliocentric radial velocity $v_{\rm hel}^{\rm c}$. Firstly, 
coordinates $x$ and $y$ are defined in the sky plane, with their axes along the apparent major and minor
axis, respectively. The corresponding coordinates in the galactic plane are $\xi=x$ and $\eta=y/\cos i$, 
where $i$ ist the galactic disk's angle of inclination as measured from face-on. Galactocentric distances 
in the sky plane are $\rho=\sqrt{x^2+y^2}$ and correspond to galactocentric distances $R=\sqrt{\xi^2 + 
\eta^2}=\sqrt{x^2 + (y/\cos i)^2}$ in the plane of the galaxy. A polar angle $\Theta$ in the plane of 
the galaxy ($\cos \Theta = \xi/R$) corresponds to $\theta$ in the sky plane via $\cos \Theta = \cos i / 
\sqrt{(\cos i)^2 + (\tan \theta)^2}$. 

To obtain the observed line-of-sight velocity component $v^{\rm los}$ of a circular (rotational) velocity 
$v^{\rm rot}$ in the galactic plane, one first projects $v^{\rm rot}$ with $\Theta$ onto a direction 
parallel to the minor axis of the galaxy and then with $i$ onto the l.o.s. so that $v^{\rm los}= v^{\rm rot} 
\, \cos \Theta\, \sin i $ or, with the above relation between $\Theta$ and $\theta$ (and $v^{\rm los} = 
v_{\rm hel}^{\rm obs} - v_{\rm hel}^{\rm c})$ :

\begin{equation}
\label{rot}
v^{\rm rot} = (v_{\rm hel}^{\rm obs} - v_{\rm hel}^{\rm c}) 
\frac {\sqrt{(\cos i)^2  + (\tan \theta)^2 }} {\sin i \cos i}
\end{equation}
%
\begin{acknowledgements}
We wish to thank an anonymous referee, M. Kadler, and D. Graham for helpful comments.
C.H. acknowledges support by NATO grant SA.5--2--05 (GRG.960086) 318/96. 
\end{acknowledgements}


\begin{thebibliography}{}
\bibitem{} Aaronson, M., Bothun, G.D., Cornell, M.E., et al. 1989, ApJ 338, 654
\bibitem{} Antonucci, R. R, J. 1993, ARA\&A 31, 473
\bibitem{} Antonucci, R. R. J. \& Miller, J. S. 1985, ApJ 297, 621
\bibitem{} Awaki, H., Ueno, S., Taniguchi, Y. \& Weaver, K. A. 2000, ApJ 542, 175
\bibitem{} Barth, A. J., Filippenko, A. V. \& Moran E.C. 1999, ApJL 515, L61
\bibitem{} Bertola, F., Pizzella, A., Persic, M. \& Salucci, P. 1993, ApJ 416, L45
\bibitem{} Bowen, I. S. 1960, ApJ 132, 1
\bibitem{} Braatz, J. A., Wilson, A. S. \& Henkel, C. 1994, ApJ 437, L99
\bibitem{} Braatz, J. A., Wilson, A. S. \& Henkel, C. 1996, ApJS 106, 51
\bibitem{} Braatz, J. A., Wilson, A. S. \& Henkel, C. 1997, ApJS 110, 321
\bibitem{} Cecil, G. 1988, ApJ 329, 38
\bibitem{} Cecil, G., Greenhill L.J., DePree C.G., et al. 2000, ApJ 536, 675 
\bibitem{} Claussen, M. J. \& Lo, K.-Y. 1986, ApJ 308, 592
\bibitem{} Claussen, M. J., Heiligman, G. M. \& Lo, K.-Y. 1984, Nat 310, 298
\bibitem{} Claussen, M. J., Diamond, P. J., Braatz, J. A., Wilson, A. S., Henkel, C. 
           1998, ApJ 500, L129
\bibitem{} Colbert, E. J. M., Baum, S. A., O'Dea, C. P. \& Veilleux, S. 1997, ApJS 105, 75
\bibitem{} Davies, R. L. \& Illingworth, G. D. 1986, ApJ 302, 234
\bibitem{} de Vaucouleurs, G., de Vaucouleurs, A., Corwin, H. G., et al. 1991, 
           Third Reference Catalogue of Bright Galaxies (RC3) (New York: Springer-Verlag)
\bibitem{} Desch, S. J., Wallin, B. K. \& Watson, W.D. 1998, ApJ 496, 775
\bibitem{} dos Santos, P. M. \& L{\'e}pine, J. R. D. 1979, Nat 278, 34
\bibitem{} Elitzur, M. 1992, ARA\&A 30, 75
\bibitem{} Elitzur, M. 1995, RevMexAA 1, 85
\bibitem{} Elvis, M. 2000, ApJ 545, 63
\bibitem{} Fairall, A. P. 1986, MNRAS 218, 453
\bibitem{} Falcke, H., Wilson, A. S. \& Simpson, S. 1998, ApJ 502, 199
\bibitem{} Falcke, H., Henkel, C., Peck A. B., et al. 2000, A\&A 358, L17
\bibitem{} Ferruit, P., Wilson, A. S. \& Mulchaey, J. 2000, ApJS 128, 139 (FWM2000)
\bibitem{} Filippenko, A. V. \& Halpern, J. P. 1984, ApJ 285, 458
\bibitem{} Fosbury, R. A. E., Mebold, U., Goss, W. M. \& Dopita M. D. 1978, MNRAS 183, 549
\bibitem{} Gallimore, J. F., Henkel, C., Baum, S. A., et al. 2001, ApJ 556, 694
\bibitem{} Gardner, F. F. \& Whiteoak, J. B. 1982, MNRAS 201, 13p
\bibitem{} Greenhill, L. J., Henkel, C., Becker, R., Wilson, T. L. \& Wouterloot, J. G. A. 
           1995a, A\&A 304, 21
\bibitem{} Greenhill, L. J., Jiang, D. R., Moran, J. M., et al. 1995b, ApJ 440, 619 
\bibitem{} Greenhill, L. J., Gwinn, C. R., Antonucci, R. \& Barvainis, R. 1996, ApJ 472, L21
\bibitem{} Greenhill, L. J., Herrnstein, J. R., Moran J. M., Menten, K.M., Velusamy, T.
           1997a, ApJ 486, L15
\bibitem{} Greenhill, L. J., Moran, J. M. \& Herrnstein J. R. 1997b, ApJ 481, L23
\bibitem{} Greenhill, L. J., Ellingsen, S. P., Norris R. P., et al. 2002, ApJ 565, 836
\bibitem{} Greenhill, L. J., Kondratko, P.T., Lovell, E.J., et al. 2003, ApJ, in press 
           (astro-ph/0212038)
\bibitem{} Hagiwara, Y., Kohno, K., Kawabe, R. \& Nakai, N. 1997, PASJ 49, 171
\bibitem{} Hagiwara, Y., Henkel, C., Menten, K. M. \& Nakai, N. 2001, ApJ 560, L37
\bibitem{} Hagiwara, Y., Diamond, P. J. \& Miyoshi, M. 2002, A\&A 383, 65
\bibitem{} Haschick, A. D. \& Baan W. A. 1985, Nat 314, 144
\bibitem{} Heisler, C. A. \& Vader J. P. 1994, AJ 107, 35
\bibitem{} Henkel, C., G\"usten, R., Downes, D., et al. 1984, A\&A 141, L1
\bibitem{} Henkel, C., Braatz, J. A., Greenhill, L. J. \& Wilson, A. S., 2002, A\&A 394, L23
\bibitem{} Herrnstein, J. R., Moran, J. M., Greenhill, L. J., et al. 1999, Nat 400, 539
\bibitem{} Ho, L. C., Filippenko, A. V., Sargent, W. L. W. \& Peng, C. Y. 1997, ApJS 112, 391
\bibitem{} Ishihara, Y., Nakai, N., Iyomoto, N., et al. 2001, PASJ 53, 215
\bibitem{} Koekemoer, A. M., Henkel, C., Greenhill, L. J., et al. 1995, Nat 378, 697
\bibitem{} Komossa, S. \& Schulz, H. 1997, A\&A 323, 31
\bibitem{} Kotanyi, C. G. \& Ekers, R. D. 1979, A\&A 73, L1
\bibitem{} Krolik, J. H. 1999, Active Galactic Nuclei: From the Central Black Hole to the
           Galactic Environment (Princeton, NJ: Princeton University Press)
\bibitem{} Lequeux, J. 1983, A\&A 125, 394
\bibitem{} Malkan, M. A., Gorjian, V. \& Tam R. 1998, ApJS 117, 25
\bibitem{} Maoz, E. \& McKee, C.F. 1998, ApJ 494, 218
\bibitem{} Martini, P. 2001, ASP Conf. Proc. 249, The Central Kiloparsec of Starbursts
           and AGN: The La Palma Connection, eds. J.H. Knapen et al. (San Francisco), p98
\bibitem{} Martini, P. \& Pogge R.W. 1999, AJ 118, 2646
\bibitem{} Mathewson, D. S., Ford, V. L. \& Buchhorn, M. 1992, ApJS 81, 413
\bibitem{} Mathis, J. S. 1970, ApJ 162, 761
\bibitem{} Melnick, G. J., Ashby, M. L. N., Plume, R., et al. 2000, ApJ 539, L87
\bibitem{} Miyoshi, M., Moran, J. M., Herrnstein, J. R., et al. 1995, Nat 373, 127
\bibitem{} M{\"o}llenhoff, C., Hummel, E. \& Bender, R. 1992, A\&A 255, 35
\bibitem{} Nagar, N. M., Oliva, E., Marconi, A., Maiolino, R. 2002, A\&A 391, L21
\bibitem{} Nelson, C. H. \& Whittle, M. 1996, ApJ 465, 96
\bibitem{} Neufeld, D. A., Maloney, P.R. \& Conger, S. 1994, ApJ 436, L127
\bibitem{} Oke, J. B. 1974, ApJS 27, 21
\bibitem{} Osterbrock, D. E. 1989, Astrophysics of Gaseous Nebulae and Active Galactic
           Nuclei (Mill Valley, CA: University Science Books)
\bibitem{} Peck, A. B., Falcke, H., Henkel, C. \& Menten, K. M. 2001, ASP Conf. Proc. 249,
           The Central Kiloparsec of Starbursts and AGN, eds. J.H. Knapen et al.
           (San Francisco), p321
\bibitem{} Peck, A. B., Tarchi, A., Henkel, C., et al. 2003, A\&A, in preparation
\bibitem{} Pelat, D., Alloin, D. \& Fosbury, R. A. E. 1981, MNRAS 195, 787
\bibitem{} Reunanen, J., Kotilainen, J. K. \& Prieto, M. A. 2002, MNRAS 331, 154  
\bibitem{} Rossa, J., Dietrich, M. \& Wagner, S. J. 2000, A\&A 362, 501
\bibitem{} Rybicki, G. B. \& Lightman, A. P. 1979, Radiative Processes in Astrophysics (New York: John Wiley \& Sons)
\bibitem{} Schmitt, H. R., Storchi-Bergmann, T. \& Fernandes, R. C. 1999, MNRAS 303, 173
\bibitem{} Schulz, H. 1987, A\&A 178, 7
\bibitem{} Schulz, H. \& Fritsch, Ch. 1994, A\&A 291, 713
\bibitem{} Schulz, H., M{\"u}cke, A., Boer, B., Dresen, M. \& Schmidt-Kaler, Th. 1995, A\&AS 109, 523 (SM95)
\bibitem{} Schulz, H., Komossa, St., Schmitz, C. \& M{\"u}cke, A. 1999, A\&A 346, 764
\bibitem{} Shields, G. A. \& McKee, C. F. 1981, ApJ 246, L57
\bibitem{} Shu, F. H. 1992, The Physics of Astrophysics II. Gas Dynamics (Mill Valley, 
           CA: University Science Books)
\bibitem{} Storchi-Bergmann, T., Fernandes, R. C. \& Schmitt H.R. 1998, ApJ 501, 94
\bibitem{} Tran, H. D. 1995, ApJ, 440, 578
\bibitem{} Trotter, A. S., Greenhill, L. J., Moran J. M. et al. 1998, ApJ 495, 740
\bibitem{} Tsvetanov, Z. I. \& Petrosian, A. R. 1995, ApJS 101, 287 
\bibitem{} T{\"u}g, H. 1977, ESO Messenger No. 11, p7
\bibitem{} Tully, R. B. 1988, Nearby Galaxies Catalog (Cambridge: CUP)
\bibitem{} Ulvestad, J. S. \& Wilson, A. S. 1984, ApJ 285, 439
\bibitem{} Veilleux, S. \& Osterbrock, D. E. 1987, ApJS 63, 295 (VO)
\bibitem{} Vrtilek, J. M. \& Carleton, N. P. 1985, ApJ 294, 106
\bibitem{} Weaver, K. A., Wilson, A. S. \& Baldwin, J.A. 1991, ApJ 366, 50
\bibitem{} Weaver, K. A., Wilson, A. S., Henkel, C. \& Braatz, J. A. 1999, ApJ 520, 130
\bibitem{} Weymann, R. J. 1970, ApJ 160, 31
\bibitem{} Whittle, M. 1985a, MNRAS 213, 1
\bibitem{} Whittle, M. 1985b, MNRAS 216, 817
\bibitem{} Wrobel, J. M. 1984, ApJ 284, 531  
\end{thebibliography}
\end{document}